\documentclass[10pt,journal,twocolumn]{IEEEtran}
\usepackage[utf8]{inputenc}
\usepackage{cite}
\usepackage{epsfig}
\usepackage{times}
\usepackage{graphicx}
\usepackage[figuresright]{rotating}
\usepackage{pict2e}
\usepackage{algorithmic}
\usepackage{algorithm}
\usepackage{amssymb}
\usepackage{amsmath}
\usepackage{setspace}
\usepackage{comment}
\usepackage{balance}
\usepackage{rotating}
\usepackage{multirow}
\usepackage{wrapfig}
\usepackage{mathtools}
\usepackage{booktabs}
\usepackage{gensymb}
\usepackage{array}
\usepackage{amssymb,amsmath,amsthm}
\usepackage[font=footnotesize]{caption}
\usepackage[labelformat=simple]{subcaption}
\usepackage{bm}

\begin{document}

\title{UW-SVC: Scalable Video Coding Transmission for In-network Underwater Imagery Analysis}
\author{{Mehdi~Rahmati and Dario~Pompili\\
{Department of Electrical and Computer Engineering, Rutgers University--New Brunswick, NJ, USA}\\
Emails: \{mehdi.rahmati, pompili\}@rutgers.edu}
\thanks{\IEEEcompsocthanksitem 2019 IEEE. Personal use of this material is permitted. Permission from IEEE must be obtained for all other uses, in any current or future media, including reprinting/republishing this material for advertising or promotional purposes, creating new collective works, for resale or redistribution to servers or lists, or reuse of any copyrighted component of this work in other works. \protect}}
\maketitle

%%%%%%%%%%%%%%%%%%%%%%%%%
%\IEEEdisplaynotcompsoctitleabstractindextext
%\IEEEdisplaynontitleabstractindextext
%\IEEEpeerreviewmaketitle

%%%pagenumbering
%\thispagestyle{empty}\pagestyle{plain}
%\pagenumbering{arabic}

\begin{abstract}
Underwater imagery has enabled numerous civilian applications in various domains, ranging from academia to industry, and from industrial surveillance and maintenance to environmental protection and behavior of marine creatures studies. 
The accumulation of litter and plastic debris at the seafloor and the bottom of rivers are extremely harmful for the aquatic life. We propose a solution for monitoring this problem using a team of Autonomous Underwater Vehicles~(AUVs) to exchange the recorded video in order to reconstruct the map of regions of interest.
%, i.e., to match the features of litter patches. 
However, underwater video transmission is a challenge in the harsh environment in which radio-frequency waves are absorbed for distances above a few tens of meters, optical waves require narrow laser beams and suffer from scattering and ocean wave motion, and acoustic waves---while long range---provide a very low bandwidth and unreliable channel for communication. In our solution, the scalable coded video of each vehicle is shared in-network with a selected group of receiving vehicles %pseudo-multicasting, 
through the underwater acoustic channel.
%The video is exchanged in-network between AUVs via sharing/merging partial information/observations of each vehicle.
%3D reconstruction can be performed to allow for efficient navigation and execution of missions to capture and reconstruct the shape and appearance of underwater objects.
%\rev{Vehicles with different capabilities can be served by a single scalable stream with different scalabilties to perform in-group 3D reconstruction.} 
Presented evaluations, including both simulations and experiments, confirm the efficiency and flexibility of the proposed solution using acoustic software-defined modems. 
\end{abstract}
\begin{IEEEkeywords}
Underwater Networks; Acoustic Communications; Broadcasting; Scalable Video Coding~(SVC).
%Software Defined Acoustic Radios;
\end{IEEEkeywords}
%%%%%%%%%%%%%%%%%%%%%%%%%%%%%%%%%%%%%%%%%%%%%%%%%

\section{Introduction}\label{sec:intro}

\textbf{Overview:}
Marine litter and debris, including both beached and floating objects, is one of the most serious and fast growing environmental threats in the oceans and seafloors. The negative impacts of litter accumulation on the aquatic life are unquestionable. Litter is spread widely throughout the seafloor, but its distribution is usually patchy with densities from $1$ item up to around $200$ items per each $10~\rm{m}$, as reported for Messina Strait's channels (one of geologically active areas of the Central Mediterranean Sea)~\cite{pierdomenico2019massive}. Rivers are one of the main sources of entering litter to the seas, since they carry the litter with their currents to the sea or ocean. Deploying a team of Autonomous Underwater Vehicles~(AUVs), equipped with down-looking cameras, can help in detecting these objects on the seafloor and riverbed, build a map of the pollution, and therefore, can issue early warnings so to reduce the damage to human and aquatic life. However, coordination among multiple AUVs is a challenge~\cite{rahmati2018PSDMA}, specially when video is the subject of data exchange. AUVs should be able to encode the video, and to transmit it to other vehicles (generally to heterogeneous dynamic nodes) efficiently~\cite{pompili2010multimedia}. There are still open problems in near-real-time underwater video processing and transmission.

To achieve these goals, novel efficient mechanisms and hardware should be utilized to make the video transmission feasible for underwater scenarios. Boosting the data rate and system reliability is possible if all the available domains are exploited in an efficient manner~\cite{rahmati2017ssfb}. To stream and transmit underwater video, we require reliable and robust techniques in an environment, in which Radio Frequency~(RF) waves are absorbed for distances above a few tens of meters, optical waves require narrow laser beams and suffer from scattering and ocean wave motions, and acoustic waves---while being able to propagate up to several tens of kilometers---lead to a communication channel that is very dynamic, prone to fading, spectrum limited with passband bandwidths of only a few tens of $\rm{kHz}$ due to high transmission loss at frequencies above $50~\rm{kHz}$, and affected by the colored ambient noise.
%~\cite{stojanovic2009underwater}.
%~\cite{akyildiz2005underwater}.

%Underwater Acoustic Networks~(UAN) enable new applications such as multimedia coastal and tactical surveillance, undersea explorations, image/video acquisition and classification, and disaster prevention, just to name a few. They should be able to retrieve multimedia data from heterogeneous nodes, to process, and to fuse them while they are being transmitted~\cite{pompili2010multimedia}. 

%Scientists estimate that every year more than $6.4$ million tonnes of litter are thrown in the oceans~\cite{pham2014marine}.

\textbf{Motivation:}
Traditional commercial acoustic modems
%~\cite{freitag2005whoi} 
with their fixed-hardware designs hardly meet the required data-rate and flexibility to support the futuristic underwater multimedia applications. Over the past few years, novel solutions based on adaptive and reconfigurable architectures---i.e., Software Defined Acoustic Radios~(SDAR)---have been proposed.
%in both the industrial research groups
%~\cite{potter2014software} 
%and the academia.
%~\cite{rahmati2018adaptive}.
%demirors2018high
Using SDAR helps the scientists and engineers to explore different protocols and techniques on a single hardware, perform in-network analysis, and transmit the high-volume data, such as video, to a remote node depending on environment and system specifications. This concept is changing the business model of commercial acoustic modems in a near future since they are focusing more on efficient hardware/architectures and proprietary high-performance algorithms~\cite{dol2017software}.

%There are many applications that benefit from this technology for video recording and transmission. Currently, archaeology in deep water is very costly and complicated using the conventional methods~\cite{singh2000imaging} with the help of divers and hand-held devices. Videos can be used in biological research and industry, e.g., for monitoring multispecies on commercial longline vessels to supplement and complement the data collected by fisheries observers~\cite{ames2007evaluation}. In many military applications, such as in the mine detection, Remotely Operated Vehicles~(ROVs) are used which are equipped with a sonar and a camera. Other applications include, but are not limited to mapping, scientific research~\cite{experimentationBruno}, environmental protection~\cite{structureBeall}, and infrastructure inspection. 

%Using the semi-autonomous vehicles require the human-in-the-loop features to enhance robustness and user Quality of Experience~(QoE) for decision making, while the futuristic full-autonomous vehicles will have enough artificial intelligence to decide on their own. 
%Furthermore, it is desirable to cut the tether and to transmit the video wirelessly.  
%\textbf{Motivation:} 

%There is an increasing need for geometric 3D models in many underwater applications due to the depth of picture delivered by this technique. 

%This makes the Quality of Service~(QoS) delivery of multimedia data a challenging task for an underwater channel.
Furthermore, using conventional video compression/encoding techniques will not meet the requirements for these futuristic underwater video transmissions due to the need for higher data rate and more reliability. Therefore, more reconfigurable and flexible techniques should be utilized to address this problem. 
%The goal of this paper is to detect the seafloor litter items can be achieved by removing some parts of the video in order to adapt it to the preferences of end users as well as to the varying condition of the communication channel to have the maximum quality of video. In practice, Simultaneous Localization and Mapping~(SLAM) can be used in autonomous missions by exploiting some features from the environment so as to create and update the map of an unknown underwater environment~\cite{durrant2006simultaneous}.
%
In practice and in many underwater imagery/streaming applications, since the visual depth of the camera is limited in the water, the vehicle should get close enough to the target to be able to detect it, therefore, usually a single vehicle/camera can not cover the whole scene (because of the limitation in the field of view and visual depth) and can not create the global map of the environment. We will also address the challenge of coordination among the underwater vehicles in this paper.

\begin{figure}[!t]
\centering
\hspace{-2mm}
\includegraphics[width=1.02\columnwidth]{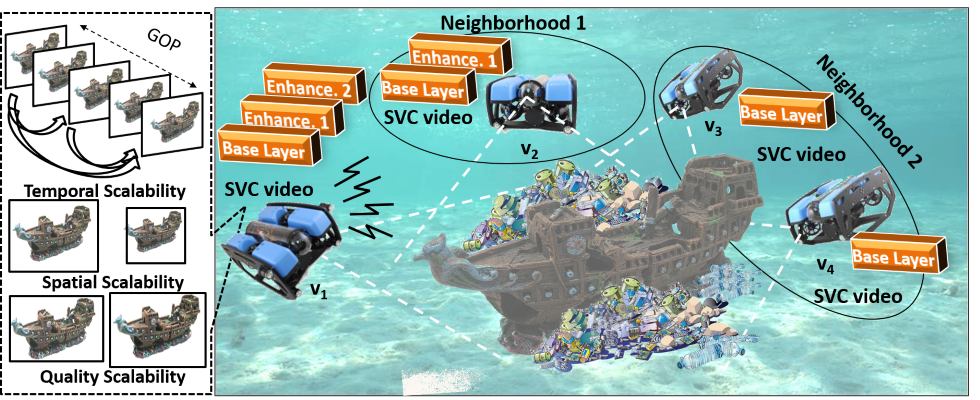}
%\vspace{-4mm}
\caption{System model of the proposed Scalable Video Coding~(SVC)-based video transmission among a team of underwater vehicles %Assume a shipwreck site requires accurate graphical representations of reality 
(with the help of high-performance modified vehicles, BlueROV2~\cite{bluerobotics}) which are used for video capturing in marine litter detection missions. Video is encoded via a base-quality layer and $l$ enhancement layers and is shared separately with each neighborhood using temporal, spatial, and quality scalabilities in SVC.}
%\vspace{-1mm}
\label{fig:sysmodel}
\end{figure}

\textbf{Our Vision:} 
We propose a solution to encode and share the video among AUVs until the global information/reconstruction of the region of interest is achieved. Scalable Video Coding~(SVC)~\cite{schwarz2007overview}, as the extension of H.264/MPEG-4 AVC, offers the required flexibility by encoding the chunks of video into a base layer and multiple enhancement layers given the requirements of the underwater channel. 
%SVC is being used in many domains during the past decade by providing the required scalabilities.
%
Fig.~\ref{fig:sysmodel} shows our vision including multiple vehicles around a pile of objects. SVC base layer provides the minimum required quality, while enhancement layers offer a more enhanced quality based on different modalities—--temporal scalability~(frame rate), spatial scalability~(frame size), and quality scalability~(fidelity or SNR)—--which makes this encoding a good choice for lossy video compression and erroneous transmission environments such as underwater. Here, a group of independent frames in the video structure is represented by a Group of Pictures~(GOP) in the figure.
%Capturing big objects (such as a pile of seafloor litter) needs the AUVs to move, 
Efficient video coding and reliable communications solutions are demanded for the coordination and communications among the vehicles. The reconstructed map can be used for in-network decision among the vehicles or
can be transmitted to the buoy for further considerations.

\textbf{Our Contributions:}
In many applications, more than one vehicle, due to the limited field of view and the visual depth of camera in the water, are needed to merge the video from different angles so as to reconstruct the map of region of interest. In this paper, we focus on in-network \textit{scalable} underwater video sharing between AUVs and offer these contributions: 
%object detection and map reconstruction of marine litter using a team of AUVs and offer the following contributions: 
\begin{itemize}
  %\item A two-step reconstruction using SVC and Structure from Motion~(SfM) based on the required accuracy and the communications constraints;
  %\item An adaptive protocol for an in-network 3D reconstruction (sparse and dense) to maximize the possibility of an accurate reconstruction after selecting the most qualified vehicle;
  \item A framework for underwater imagery analysis using partial information collected by various vehicles around the scene;
  \item An optimized solution to provide the maximum possible Quality of Service~(QoS) via a proposed multicasting scalable coded video, while achieving the maximum Quality of Experience~(QoE) for the scene reconstruction;
\item Performance evaluation of this system with comprehensive simulations under different scenarios using real videos captured from the Raritan river-New Jersey and through an SDAR testbed.
\end{itemize}

\textbf{Paper Organization:}
%The remainder of the paper is organized as follows. 
In Sect.~\ref{sec:rel_work}, we go over the state of the art in underwater video transmission. In Sect.~\ref{sec:prop_soln}, we present our solution and discuss scalable video coding and the required optimizations.
%; then, a novel protocol for data exchange among the vehicles is presented. 
In Sect.~\ref{sec:Eval}, we evaluate our solution via the experiments and simulations, and then scale the results via simulations. %and discuss the benefits of implementing our solution. 
Finally, in Sect.~\ref{sec:Con}, we draw the main conclusions and present the future work.

%%%%%%%%%%%%%%%%%%%%%%%%%%%%%%%%%%%%%%%%%
\section{Related Work}\label{sec:rel_work}

%\todo{you can have two bold headings to cluster the related works, one on video and one on uw video comms}

%\todo{also, make sure it's clear how our work positions w.r.t. related work}

\textbf{Underwater Video Transmission:}
There are several unique characteristics of underwater wireless networks that make Quality of Service~(QoS) delivery of video content---ranging from delay sensitive to delay tolerant, and from loss sensitive to loss tolerant---a challenging task due to underwater acoustic frequency-dependent transmission loss, colored noise, multipath, Doppler frequency spread, high propagation delay as discussed in~\cite{pompili2010multimedia,stojanovic2007relationship}. The multiview video transmission in underwater acoustic path is discussed in~\cite{fujihashi2018multi} in which the authors propose time-shifted transmission slots to the encoder and other nodes to exchange control and video packets. The feasibility of transmitting video over short-length underwater links is investigated in~\cite{ribas2010underwater,vall2011towards}, where MPEG-4 video compression and a wavelet-based transmission method are tested on the coded Orthogonal Frequency Division Multiplexing~(OFDM). Despite all these works, the problem of robust video transmission is still unsolved, and achieving high video quality is still a challenge when we consider the limited available bandwidth along with the harsh characteristics of the underwater acoustic channel, \textit{which calls for novel high-spectral-efficiency in-network collaborative methods.} In the area of underwater video, \cite{underwaterFeasbility} shows the feasibility of video streaming using currently commercially available hardware defined modems. The reconstructed objects can be used in  Simultaneous Localization And Mapping~(SLAM). SLAM is a widely used technique in ground robots, but less feasible in underwater environment specially in high turbidity situations and in the absence of reliable static landmarks. Some underwater visual SLAM solutions, such as in~\cite{rahmati2018slam}, create a sparse map for the navigation and localization in clear water. 

\textbf{Scalable Video Coding~(SVC):}
SVC~\cite{schwarz2007overview} outperforms the regular H.264 encoding when more flexibility and adaptation to the channel's condition are required~\cite{stutz2012survey}. In the area of SVC, previous papers have touched on video sharing/multicasting in terrestrial context.
%\cite{AdaptiveSVCWireless,P2PSVC,linkAdaptiveScheme}. 
A method for adapting the number of layers based on a fixed time allotment is proposed in~\cite{linkAdaptiveScheme}, This link-level method does not explore a multicast scenario. The authors in~\cite{P2PSVC} explore dynamic layer adjustment in a content-delivery context where a direct-download system is paired with peer-to-peer. This sharing is top-down content delivery, rather than a scheme for in-group video sharing where each consumer is also a producer. A method for SVC video transmission is proposed in~\cite{SVCWireless} using transmitter-side distortion estimates based on the channel state information. However, \textit{none of these methods tackle the unique challenges faced in an underwater acoustic channel.}

%
%\rev{more on multi-vehicle SLAM even in the air}
%\rev{
%1-a review on important SDARs in other groups and their latest results (including ours) (you can take some references from ~\cite{dol2017software}\\ 
%Adam! please add some new/classic trends on multimedia encodings (ONLY IN UNDERWATER). some new trends on SVC (both air and underwater if any to show it has merit). some new trends in 3D Reconstruction (both air and underwater. sparse and dense reconstructions. any other related work}
%\rev{4- some new trends in multicast video transmission}
%5- 3D Reconstruction (Structure from Motion, SLAM)?\\
%\reword{M: yes, we can. SLAM is basically a different story, I would say not involve it here in this paper, just mention a couple of lines n the related work. Please cite our paper on SLAM as well, as follows.}

An adaptive distortion-rate tradeoff for underwater video transmission using a Multi-input Multi-output~(MIMO)-based SDAR system is proposed in~\cite{rahmati2018adaptive}. The scalability of the system is fulfilled using SVC compression standard. In~\cite{rahmati2017ssfb} a new signaling for SVC-encoded underwater videos is proposed based on using non-contiguous OFDM and beamforming techniques with the help of Acoustic Vector Sensors~(AVSs).

\section{Our Solution}\label{sec:prop_soln}
In this section, we present our solution for in-network video sharing and coordination among multiple AUVs. In Sect.~\ref{subsec:svc}, we discuss the construction of SVC-encoded video streams and the proposed strategy to estimate the optimal parameters given underwater acoustic channel constraints as it will be explained in the optimization problems. 
In Sect.~\ref{subsec:multi},
we present our SVC-based multicasting solution to increase the overall quality of video. In Sect.~\ref{subsec:consen}, the proposed protocol will be presented for an efficient map reconstruction while multiple vehicles are involved in the merging process.

%In Sect.~\ref{subsec:protocol},
%We present our protocol and the required steps we take for an efficient marine litter detection solution for the layered video stream. 
%In Sect.~\ref{subsec:3Ddense}, we describe the procedure for a global map reconstruction in the selected vehicle. 

\subsection{Construction of SVC-encoded Video Streams}\label{subsec:svc}

%\textbf{SVC-encoded Video:} 
Encoding the original video into several layers using SVC discards the need for transcoding or re-encoding the video. However, an efficient strategy is required to leverage the scalibiliteis of SVC and adapt the encoder to the receiver's status as well as the quality of acoustic channel. 

%Fig.~\ref{fig:svc} \rev{[to be fixed]} shows an example of the main scalability types---i.e., temporal, spatial, quality-based---in SVC, in which the group of independent frames in the video structure is represented by a Group of Pictures~(GOP). 
% %Fig.~\ref{fig:diagram} depicts the system diagram and the required actions which will be explained in details in the following sub-sections.  

\textbf{Video Sharing Setup:}
Assume $V$ vehicles are deployed  around a scene, as shown in Fig.~\ref{fig:sysmodel}, at time slot $t$ and form a wireless network of $(V,H)$, where $H$ stands for the point to point link between two vehicles, when vehicles are in the communications range of each other. Vehicles encode the initial video using SVC, and make it ready for broadcasting. To facilitate the communications, vehicles set up a basic Time Domain Multiple Access~(TDMA) system and assign a time slot to each vehicle since the network size is small in underwater scenarios and the nodes are usually close together. The underwater acoustic channel presents problems for a coordinated and synchronized system such as TDMA, but due to the severe bandwidth constraint, it is important to use a Medium Access Control~(MAC) that does not constrain vehicles to an even smaller slice of bandwidth, such as FDMA. Authors in~\cite{lmai2017throughput} show that even in the underwater acoustic environment, and specially for multicast transmissions, TDMA can allow for efficient and collision-free communications. Other random- and controlled-access MAC solutions such as Carrier-sense Multiple Access~(CSMA) transmit multiple packets through the same underwater channel, which might lead to packet collisions at the receiver~\cite{rahmati2018PSDMA}. To address the synchronization problem in TDMA (as the main weakness of using TDMA underwater), we use an unsynchronized MAC protocol, e.g., Tone Lohi~(T-Lohi)~\cite{syed2008t}, especially in sparse networks with limited number of nodes. The vehicles start contending any time they realize the channel is not occupied.  

%?%%% Once the schedule is finalized, the additional time can be used to transmit pilot signals for channel estimation between all nodes so that each vehicle can use the state of the channel to inform their encoding decision. 

%\rev{the max amount of overlap at the nodes due to the fact that we are doing scheduling in broadcast instead of talking directly to the gateway}

%\rev{how to select the node to speak up}

%One of the constraints in TDMA is how to define the minimum frame length for each vehicle. Therefore, $T_f>????$, \rev{[calculations on the min. frame time]} where $T_f$ is This depends on the propagation delays 

\textbf{Base-layer Video Sharing:}
Assume each vehicle records the scene from its own angle and possibly it has an overlapping coverage with other vehicles. SVC-based video is segmented into $C$ chunks in each vehicle $j \in \{1,...,V\}$ with a base layer $b_j$ (layer $0$) with the rate $R(b_j)$ and $l_j \in 1,2,...,L_j$ enhancement layers with rate $R(l_j)$. Each node broadcasts the chunks of its base layer video through an acoustic channel. 
%which has a low quality and low frame rate in turn. 
%This base layer, along with a pilot signal for the purpose of channel estimation, is transmitted to all other nodes. 
When a vehicle $i$ receives the base layer data of chunk $c\in C$ in time slot $t$ from transmitting vehicle $j\in\{1,...,V\}$ and $j\neq i$ in the communication range, the received signal can be expressed as $y_i^c(t)=h_{ij}^c(\tau_{ij};t)\ast b_j^c(t)+z_i(t)$, where $h_{ij}^c(\tau_{ij}^c;t)$ stands for the channel coefficient with delay $\tau_{ij}^c$ between vehicles $i$ and $j$,  $y_i^c(t)$ represents the received signal, $\ast$ stands for the convolution operation, and $z_i(t)$ shows the background underwater colored noise. For a band-limited non-ideal underwater channel with the frequency response of $H_{ij}^c(f)$ and a Gaussian noise with the power spectral density of $S_i(f)$, the capacity $\mathcal{C}$ of each channel can be expressed as~\cite{john2008digital},
\begin{equation}\label{eq:c}
    \mathcal{C}_{ij}^c=\frac{1}{2}\int_{-\infty}^{\infty}\log\Big(1+\frac{P_j^c(f){\mid H_{ij}^c(f)\mid}^2 }{S_i(f)}  \Big)df.
\end{equation}

Here, $P_j^c(f)$ stands for the power spectral density of $b_j^c$ from transmitting vehicle $j$ in chunk $c$. We drop time index $t$ for the sake of simplicity and present our analysis for the time length of chunk $c$.  
Assume Channel State Information~(CSI) is available at the transmitter and the channel is constant during broadcasting of a video stream in chunk $c$ and $B_W$ represents the channel bandwidth, which is assumed to be the same for all the users. The base layer data rate $R_{ij}(b_j^c)$ can be expressed as $R_{ij}(b_j^c)=B_W \mathcal{C}_{ij}^c$. We consider the tradeoff between the transmit power and data rate for a fixed bandwidth $B_W$ in each vehicle $j$ such that the outage does not occur. Since we assume each vehicle $j$ broadcasts its data to all other vehicles in its neighborhood through \textit{independent channels}, the broadcast data rate ${\bm{R}_{j}(b_j)}_{BC}$ for all chunks can be bounded as follows.
\begin{equation}
{\bm{R}_{j}(b_j)}_{BC}=\{{R_{ij}(b_j^c):{{R}_{m,j}^*(b_j)}<{R_{ij}(b_j^c)} <\mathbb{E}[\mathcal{C}_{ij}]}\}.
\end{equation}

%\begin{lemma}\label{lemma1}
%?? \rev{pause}
%\end{lemma}

In this equation, $\mathbb{E}[.]$ represents the expectation operator, ${\bm{R}_{j}(b_j)}_{BC}$ stands for the practical transmission rate for broadcasting, and ${{R}_{m,j}^*(b_j)} \in {\bm{R}_{j}^*(b_j)}=[{R}_{1j}^*(b_j),...,{R}_{V-1j}^*(b_j)]$ shows the minimum rate required in all fading situations\cite{jindal2003capacity} for $V-1$ receiving vehicles to avoid an outage. 

In practical scenarios, in which the CSI is not fully known at the transmitting vehicle and channel gains are not known in advance, we assume that the transmitting vehicle $j$ statistically knows the ordering of the other vehicles for each chunk $c$ in time slot $t$ in terms of their instantaneous channel gains, i.e., $\mid h_{1j}^c \mid < \mid h_{2j}^c \mid < ... <\mid h_{3j}^c \mid$, for receiving vehicles $1,..., V-1$, from weak to strong. The broadcast channel can be considered as a multiple-component channel such that a weaker component is a degraded version of the other component in a symmetric broadcast channel. It can be proved that the vehicles have the same channel quality and hence could decode the broadcast data. Here, the fading statistics are assumed to be symmetric. Considering the principle of ergodicity, 
%and similar to the case of Additive White Gaussian Noise~(AWGN), 
if an arbitrary user $k$ can decode its data reliably, then we can conclude all the other users should be able to decode the broadcast data in the same way. This assumption breaks in the asymmetric fading case in which the users have different fading distributions. Therefore, sorting is not possible which leads to a non-degraded channel~\cite[Ch.~6]{tse2005fundamentals}. 

We optimize the total rate for broadcasting from vehicle $j$ to other vehicles via the following optimization problem.
\begin{subequations}\label{opt_prob1}
        \begin{align}
         \underset{p_j}{\rm{maximize}}  \;\;\;& \mathbb{E}\bigg[\sum_{\underset {i\neq j}{i=1}}^{V}\alpha_i\log\Big(1+\frac{p_j^c{\mid h_{ij}^c\mid}^2 }{s_i}\Big)\bigg],    \label{opt_prob1_a} \\ 
         \rm{s.t.}  \;\;\;  & p_{th}\leq p_j^c \leq p_{max}, \;\;\;\;\, \forall j \in \left\{1,...,V\right\},\label{opt_prob1_b}   \\ 
         & {\bm{R}_{j}(b_j)}_{BC} \succeq  R_j^*(b_j) \bm{1} \label{opt_prob1_c},\\
         & {\bm{R}_{j}(b_j)}_{BC} \preceq \mathbb{E}[\bm{\mathcal{C}}_{j}].
         \end{align}
\end{subequations}

Here $\alpha_i \in \{0,1\}$ is the weighting factor, which is defined in the multicasting strategy, $p_{th}$ and $p_{max}$ show the minimum and maximum transmit power, respectively. $\bm{1}$ stands for an all-one vector, i.e., a vector whose entries are all equal to one, $\succeq$ and $\preceq$ represent the component-wise inequality. The capacity $\bm{\mathcal{C}}_{j}$ stands for the vector of all capacities to the receiving vehicles. 

The optimization problem presented in~\eqref{opt_prob1} is a convex problem, since the objective function and the constraints are convex/concave; $\log(1+{p_j^c{\mid h_{ij}^c\mid}^2 }/{s_i})$ is concave because it is the composition of a concave function ($\log$) with an affine mapping of $p_j^c$. Moreover, the non-negative weighted sum preserves the convexity (concavity) and the expectation of a convex (concave) function is convex (concave)~\cite{boyd2004convex}. Furthermore, the constraints are all affine.         

%%%%%%%min-rate capacity region-- check later
%\rev{The minimum-rate capacity region of a broadcast channel should be defined for $V$ vehicles as the region of all achievable average rate vectors subject to an average power constraint and minimum rate constraints~\cite{jindal2003capacity}.}

%\rev{If we assign some length of time $\tau_b$ for the base layer sharing step, no transmitter $j$ among the $N$ transmitters should take more than $\tau_b/N$ for its transmission. Therefore, if it has a rate $R_{min} = \rm{min} \{R_i ;i\}$ over the worst channel to another node, it should encode its base layer video of length $t$  at a rate $r_b \leq R_{min}\tau_b/(N t)$ in order to ensure that all nodes may share their base layer video before the allotted time $\tau_b$ elapses.}
%%%%%
%\rev{Adam, can you continue this discussion from the video encoder's side? SVC coder in vehicle $j$ should not have a base layer with a rate more than $R(b_j)$. }\blue{We are not limited to real time video transmission, we can take any amount of time to share 10s of (very low frame rate) video. Faster is better but I don't see a hard constraint}
%A: What do we get out of the limiting assumption that we can estimate channels before the base layer sharing step?

%\textbf{Multicasting Strategy for Enhancement-layer Video Sharing:}
In a broadcast scenario, each transmitting vehicle propagates its base layer video to all the receiving vehicles, since decoding the base layer is independent of other enhancement layers. However, the optimized data rate, calculated in~\eqref{opt_prob1}, might not be sufficient for a higher quality video through the enhancement layers. Each enhancement layer $l_j$ with a defined encoding rate of $R(l_j)$ can be decoded when firstly it is received reliably and secondly its lower layer $l_j-1$ is successfully decoded, i.e., in other words, unsuccessful decoding of the lower layers leads to a failure in decoding the current layer.   

\subsection{Multicasting for Enhancement-layer Video Sharing}\label{subsec:multi}
In a multicast scenario and due to heterogeneity of underwater nodes, we assume the nodes with poor channel quality are able to decode the video with the base layer (as discussed in Sect.~\ref{subsec:svc}), while the nodes with a better communications channel quality can be served by a scalable video with a higher quality, i.e., with more enhancement layers. 
%The base layer has the minimum size and is a valid H.264 video with any arbitrary parameters.  Enhancement layers can, when added to the base layer, provide additional benefits in different dimensions, i.e., temporal, in which enhancement layers increase frame rate; spatial, where they increase resolution; and quality, where they increase the Quantization Parameter~(QP). Quality scalability also can be interpreted as a special case of spatial scalability with identical picture sizes in base and enhancement layer~\cite{schwarz2007overview}. 
To be able to send the enhancement layers, we propose a broadcasting strategy in which the vehicles with the worst channel are shut down in the broadcasting, i.e., $\alpha_i=0$ in~\eqref{opt_prob1_a}, in order to increase the total transmission data rate. Therefore, a pseudo-multicasting network is created. Apparently, the more vehicles with impaired channels are shut down, the more enhancement layers can be transmitted to the remaining vehicles and therefore video QoS increases. 

\begin{figure}[!t]
\centering
\hspace{-2mm}
\includegraphics[width=0.95\columnwidth]{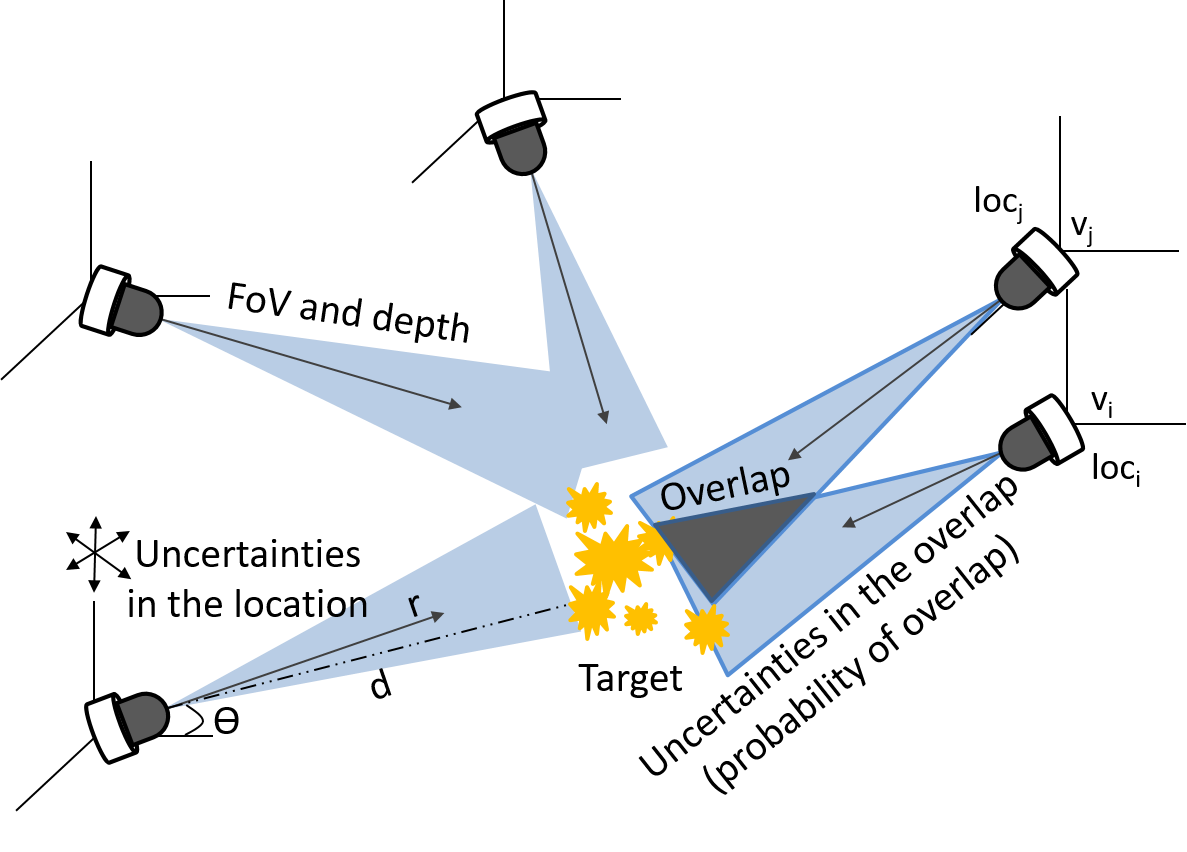}
%\vspace{-4mm}
\caption{Schematics of the potential overlap between the vehicles considering the uncertainties in the location of vehicles.}
%\vspace{-2mm}
\label{fig:overlap}
\end{figure}

On the other hand, since the vehicles are at different locations around the scene with different viewpoints (as it is depicted in Fig.~\ref{fig:overlap}), shutting them down, leads to lack of observation and so it results in losing some information while the map is reconstructed. Map reconstruction requires a good amount of Fields of View~(FoV) overlap among the vehicles.
Assume the vehicles' cameras have some degrees of spatial correlation, as shown in Fig.~\ref{fig:overlap}, which is identified via the vehicles' configuration, i.e., area of overlap between FoVs of two cameras~\cite{pandey2018robust}. The FoV of cameras is limited to the area they observe, therefore, the information they get is directly related to the directional sensing and configuration of the vehicle. This overlap is used by the algorithm as a measure to shutdown the redundant vehicles if there exists a sufficient overlap for map reconstruction. 

Let the FoV model of vehicle $i$, after 3-D to 2-D projection and calibration, be described by $(loc_i,r_i,\vec{D_i},\beta_i)$ as in~\cite{dai2009spatial}, in which $loc_i$ stands for the location of the vehicle, $r_i$ represents the sensing radius of the camera, $\vec{D_i}$ indicates the sensing direction (i.e., the center line of sight of the camera's FoV), and $\beta_i$ is the offset angle. 
%Focal length of each camera can be estimated by $f_i$ as shown in~\cite{devarajan2008calibrating}. 
A model for the spatial correlation can be derived based on the above parameters as follows. Suppose vehicles $i$ and $j$ are two arbitrary vehicles that observe an overlapped area of interest; their disparity function $\delta$  (complementary to the correlation coefficient $\eta$ as $\delta=1-\eta$) is defined as follows~\cite{dai2009spatial}: 
$\delta = {\frac{1}{4}} \Big( \left | \frac{d \sin \theta}{d+\cos\theta} \right | + \left | \frac{d \sin \theta}{d-\cos\theta} \right |+ \left | \frac{d \cos \theta}{d+\sin\theta} -1\right |+\left | \frac{-d \cos \theta}{d-\sin\theta} +1\right | \Big)$, 
%\end{split}
%\end{equation}
\begin{comment}
\begin{equation}
\scalebox{0.94}[1]
\begin{split}
&\delta = {\frac{1}{4}} \Bigg( \left | \frac{d \sin \theta}{d+\cos\theta} \right | + \left | \frac{d \sin \theta}{d-\cos\theta} \right |+ \left | \frac{d \cos \theta}{d+\sin\theta} -1\right |\\
&+\left | \frac{-d \cos \theta}{d-\sin\theta} +1\right | \Bigg), 
\end{split}
\end{equation}
\end{comment}
where $d$ denotes the camera depth (here, the difference between the $loc_i$ and the target's location assuming the camera sensing direction $\vec{D_i}$ is headed to the target) and $\theta$ is the angle between the sensing direction and the $x$-axis, so that the location $loc_i$ can be expressed by $(-d \cos \theta, -d \sin \theta)$ after the 2-D projection. Specifically, for two vehicles~$i$ and $j$ with parameters $(d_i,r_i,\theta_i)$ and $(d_j,r_j,\theta_j)$, respectively, the disparity between their images can be calculated as follows~\cite{dai2009spatial,pandey2018robust},
${\delta _{i,j}} = {\frac{1}{4}}$
{$\Big(\left\vert{{-d_{i}\sin \theta _{i}-r_{i}\cos\theta _{i}}\over{d_{i}+\cos\theta _{i}}}- {{-d_{j}\sin \theta _{j}-r_{j}\cos\theta_{j}}\over {d_{j}+\cos \theta _{j}}}\right\vert +
\left\vert {{d_{i}\sin \theta _{i}+r_{i}\cos\theta _{i}}\over {d_{i}-\cos \theta _{i}}}- {{d_{j}\sin \theta _{j}+r_{j}\cos\theta_{j}}\over {d_{j}-\cos \theta _{j}}}\right\vert +
\left\vert {{d_{i}\cos \theta _{i}-r_{i}\sin\theta _{i}}\over {d_{i}+\sin \theta _{i}}} - {{d_{j}\cos \theta _{j}-r_{j}\sin\theta_{j}}\over {d_{j}+\sin \theta _{j}}}\right\vert + \Bigl\vert {{-d_{i}\cos \theta _{i}+r_{i}\sin\theta _{i}}\over{d_{i}-\sin \theta _{i}}}-{{-d_{j}\cos \theta _{j}+r_{j}\sin\theta_{j}}\over {d_{j}-\sin \theta _{j}}}\Big\vert \Big).$} However, finding the exact amount of correlation might not be feasible due to the position uncertainty of the vehicles and the effect of currents on the vehicles due to vehicle's drifting. Therefore, inaccuracies in position estimation increases and it becomes worse over time when the vehicle stays longer underwater, which leads to non-negligible drifts in the vehicle's position and thus making the camera overlap accurate calculations inapplicable.

In~\cite{rahmati2018PSDMA}, 
an approach has been proposed to estimate vehicles' position through a statistical method based on the vehicles' confidence region. Assume each vehicle $i$ measures $N$ random samples of its location as $\{{loc}_i^{(n)}\}_{n=1}^{N}$. The measured locations are samples of a normal distribution $\mathcal{N}(\mu_i,\sigma_i^2)$ with the mean and variance $\mu_i$ and $\sigma_i^2$, respectively. The samples also follow a normal distribution with mean $\mu'_i$ and variance ${\sigma'}_i^2$. It can be inferred that $\dfrac{\mu'_i-\mu_i}{{\sigma'}_i^2/\sqrt{N}}$ is a pivot and it has a student's t-distribution with $N-1$ degrees of freedom. The mean $\mu'_i=\sum_{n=1}^{N}{loc}_i^{(n)}/{N}$ and the variance can be estimated as ${\sigma'}_i^2={1}/{(N-1)}\sum_{n=1}^{N}\left ( {loc}_i^{(n)}-\mu'_i\right)^2$~\cite{rahmati2018PSDMA}. 
The uncertainty region, i.e., confidence interval, of this vehicle can be derived as $\Pr(L_i\leq \mu'_i \leq U_i)\geq 1-\gamma$. Here $\gamma$ is the confidence degree, $\Pr(.)$ represents the probability function, and $L_i$ and $U_i$ are the interval boundaries of vehicle $i$ and are estimated as $L_i=\mu'_i-\mathcal{T}_{(N-1,\alpha /2)}{\sigma^2}/{\sqrt{N}}$ and $U_i=\mu'_i+\mathcal{T}_{(N-1,\alpha /2)}{\sigma_i^2}/{\sqrt{N}}$. Here, $\mathcal{T}_{N-1,\alpha /2}$ is the t-distribution critical value with $N-1$ degrees of freedom. To estimate the amount of overlap between two vehicles $i$ and $j$, we define the probability of overlap as $\Pr_{i,j}^{(o)}= \Pr({\eta}_{i,j}>0)=\Pr({\delta}_{i,j}<1)$, we have,
$\rm{Pr}_{i,j}^{(o)}=\int_{0}^{\infty}f({\eta}_{i,j})d{\eta_{i,j}}=\frac{1}{\sigma _{i,j}\sqrt{2\pi }}\int_{0}^{\infty } \exp {\left \{ -\frac{1}{2}\left (\frac{{\eta}_{i,j}-\mu _{i,j}}{\sigma_{i,j}}  \right )^{2} \right \}} d{\eta}_{i,j}$.
%
\begin{comment}
\begin{equation}\label{xprim17}
\rm{Pr}_{i,j}^{(o)}=\int_{0}^{\infty}f({\eta}_{i,j})d{\eta_{i,j}}=\frac{1}{\sigma _{i,j}\sqrt{2\pi }}\int_{0}^{\infty } \exp {\left \{ -\frac{1}{2}\left (\frac{{\eta}_{i,j}-\mu _{i,j}}{\sigma_{i,j}}  \right )^{2} \right \}} d{\eta}_{i,j}.
\end{equation} 
\end{comment}
%
By defining the auxiliary variable $x={({\eta}_{i,j}-\mu _{i,j})}/{\sigma_{i,j}}$, we obtain,
\begin{equation}
\rm{Pr}_{i,j}^{(o)}=\frac{1}{\sqrt{2\pi }}\int_{-\infty}^{({\mu _{i,j}}/{\sigma_{i,j}})}e^{-({x^2}/{2})}dx=\Phi(\frac{\mu _{i,j}}{\sigma_{i,j}}),
\end{equation}
where $\Phi(.)$ is the Cumulative Distribution Function~(CDF) of the standard normal distribution. 

The following optimization problem in~\eqref{opt} justifies the discussion on the number of enhancement layers that the transmitter can handle on the top of the encoded base layer video. This is a knapsack program, which defines the enhancement layers of rate $R_j(l_j)$ that could be transmitted over the underwater channel with maximum achievable communication data rate $R_{max}$,
\begin{subequations}\label{opt}
\begin{align}
%&\mathop {\max }\limits_{R_i,\alpha_i}  \;\;  \frac{\sum_{i=1}^{L+1} \alpha_i R_i}{D_e+\sum_{i=1}^{L+1} \alpha_iD_{ci}}
%&\mathop {\rm{maximize} }\limits_{\lambda_k}  \;\;  \sum_{l=1}^{L} \lambda_l \lambda_{l-1} R_j(l_j) \label{x16} \\ 
\underset{\lambda_k}{\rm{maximize}}  \;\;\;& \sum_{l=1}^{L} \lambda_l \lambda_{l-1} R_j(l_j),    \label{opt_prob1_a} \\ 
%&\text{s.t.} \hspace{0.51cm}  D_e+\sum_{i=1}^{L+1} \alpha_iD_{ci} \leq D_T, \label{x16-1} \\
\text{s.t.} \;\;\;  &  \sum_{l=1}^{L} \lambda_l \lambda_{l-1} R_j(l_j) \leq R_{max},\label{x16-3}\\
%& \hspace{0.9cm}  R_{min} \leq \sum_{i=1}^{L+1} \alpha_i R_i \leq R_{max},\label{x16-3}\\
%& \hspace{0.9cm} R_i\geq R_{el}\\
& \lambda_0=1, \; \lambda_l \in \{0,1\}, \forall l
\in \{1,...,L\}.
\end{align}
\end{subequations}

We determine the minimum number of vehicles to shut down such that we achieve the required QoS in the received video with an acceptable Quality of Experience~(QoE) in the reconstructed map of environment based on a defined amount of spatial correlation. Vehicles are eligible to transmit a video with higher enhancement layers while the layers bellow are successfully received/decoded. In this case, the following optimization problem can be presented for every chunk $c$ of the video, given the optimal power $P_j$ and the data rate $R_j$ calculated from~\eqref{opt_prob1},
\begin{subequations}\label{opt_prob2}
\begin{align}
\underset{\alpha_i}{\text{maximize}} \;&
\sum_{i=1}^{V-1}\alpha_i \label{opt_prob2a}\\
s.t.  \;  & \alpha_i \in \{0,1\}, \label{opt_prob2c}\\
& \mathbb{E}\bigg[\sum_{\underset {i\neq j}{i=1}}^{V}\alpha_i\log\Big(1+\frac{p_j^c{\mid h_{ij}^c\mid}^2 }{s_i}\Big)\bigg] \geq {\rm{QoS}}_{th}(l_j),\label{opt_prob2d}\\
& D_i<D_{th},\label{opt_prob2e} \\
& \rm{Pr}_{i,k}^{(o)}\geq Pr_{th}, %\delta_{ii-1} < 1 , 
\;\;\;\;\, \forall i,k \in \left\{1,...,V-1\right\},  \label{opt_prob2b}
%%& \sum_{k=0}^{l_j}r_{k}\tau_v \leq C_{v,u}\tau_e
\end{align}
\end{subequations}
%%%%%%%%%%%%%%%%%%%%%%%%%%%%%%%%%%%%

where the objective function~\eqref{opt_prob2a} is the total number of vehicles. Maximizing the total number of vehicles (i.e. minimizing the number of vehicles to shut down) ensures the QoE since more vehicles from different angles are present in the map reconstruction. On the other hand, to satisfy a threshold QoS, the proposed method will shut down the vehicles with the worst channel to keep the average broadcasting rate over a minimum value, as shown in~\eqref{opt_prob2d}. The other metric for QOS is represented in constraint~\eqref{opt_prob2e} which is defined by the SVC encoder and depends on the scalability and the number of enhancement layers that the encoder uses.
%as explained in Table~\ref{tab:SVCQoS}.
For an encoded video, we can write~\cite{stuhlmuller2000analysis}
$D_i={\hat{\theta}}/{(R_j-R_0)}+D_0$,
%To transmit a video through this channel, based on the Rate-Distortion~(RD) model of a compressed video, we have~\cite{stuhlmuller2000analysis},
where $D_i$ represents the distortion of the video at the vehicle $i$ at the time of reconstruction and $R_j$ is the rate of the encoder at vehicle $j$; the other remaining variables $\hat{\theta}$, $R_0$, and $D_0$ depend on the encoded video and on the model, and are estimated empirically.
The last constraint~\eqref{opt_prob2b} shuts down the vehicles which have a higher probability of overlap with the neighboring vehicles to have the minimum reduction in the QoE in reconstruction from different angles. 

%where $r_{k}$ denotes the data rate of the video layer $k$, $\tau_v$ denotes the length of $v$'s video, and $C_{v,u}$ denotes the capacity of the channel between vehicle $v$ and the FRV $u$. The constraint is a non-negative weighted sum of a convex function and so it is convex. \rev{-revise-}

\begin{comment}
{The rest of the vehicles must use the remaining chunk time $\tau_e$ to transmit as many layers as possible. Given that we have $K$ transmitting vehicles in this step, each one has $\tau_e/K$ time to transmit. Vehicle $v$ should transmit a $L_v$ subject to the optimization problem, 
\begin{equation}
\begin{aligned}
& \underset{L_v}{\text{maximize}}
& & L_v \\
& \text{subject to}
& & \sum_{k=0}^{L_v}r_{k}\tau_v \leq C_{v,u}\tau_e
\end{aligned}
\end{equation}
 where $r_{k}$ denotes the bitrate of the video layer $k$, $\tau_v$ denotes the length of $v$'s video, and $C_{v,u}$ denotes the capacity of the channel between vehicle $v$ and the FRV $u$. 
 The constraint is a nonnegative weighted sum of a convex function and so it is convex.}
\end{comment} 

\subsection{In-network Marine Litter Map Reconstruction}\label{subsec:consen}

As it was discussed in the previous sections and due to the limited FoV of each single vehicle, a cooperation among the vehicles is required so that the required map can be reconstructed. %In this section, multiple map reconstruction strategies will be discussed. 

\begin{figure}[!t]
\centering
\hspace{-3mm}
\includegraphics[width=1.02\columnwidth]{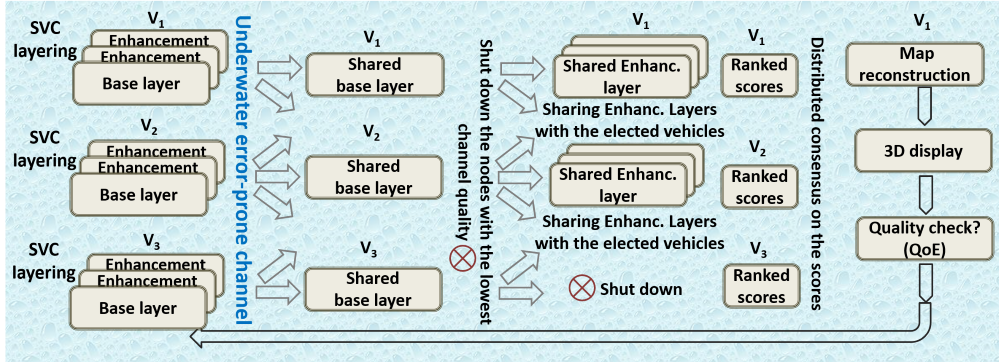}
%\vspace{-4mm}
\caption{Required steps for a satisfactory map construction. Vehicles $V_1$-$V_3$ share their encoded video (base layer) with other nodes. Enhancement layers are shared with the vehicles with a better acoustic channel quality (lower quality channels are shut down). After reaching a consensus, final reconstruction is performed on the highest-rank vehicle after it receives a high-quality video with higher QoS from other eligible nodes. If the QoE is not satisfactory, the process is restarted using a feedback command.}
%\vspace{-2mm}
\label{fig:diagram}
\end{figure}

\textbf{Potential Cooperation Strategies:}
We propose different strategies based on the exchanged data, acoustic channel requirements, level of complexity (that the vehicles can handle to process the data locally) and the QoS/QoE requirements as follows:
$(i)$~Vehicles exchange their local maps after each partial map is created. This strategy requires the minimum amount of data exchange since the merger creates the global map based on only a consensus on the exchanged local maps. 
%, what are these local maps? how do we find them? when sharing them we assume there is no overlap between them, right? how does the fusion work? is it distributed or centralized?
$(ii)$~Vehicles exchange the SVC-based channel independent videos, i.e., base layers.
%can we interpret this as the sparse reconstruction (get inspired from the previous discussion in 3D reconstruction)
%
$(iii)$~Vehicles exchange SVC adaptive channel dependent video, i.e., base and enhancement layers. This is the most desirable strategy that is also adaptive with the channel quality. 
%can we take it as dense reconstruction? then I think the discussion on the channel comes in this section.
$(iv)$~Vehicles exchange the high quality video considering the acoustic channel bandwidth and the channel fading. This strategy is usually not feasible underwater due to the bandwidth limitation and time-varying nature of the underwater acoustic channel.
%
%Maybe we can show that it is not usual/feasible unless it has a huge bandwidth and a very good channel which is optimistic.
%\subsection{Sparse Map Reconstruction}\label{subsec:obj}
%
Fig.~\ref{fig:diagram} depicts the strategy we choose in this paper. After sharing the base layer, as we discussed in the previous sections, we shut down the vehicles with unreliable channels to be able to reach the required rate for sending the enhancement layers. We lose some part of the scene from those nodes which experience the shut down. Therefore, the vehicles should reach a consensus to decide on the node who finally reconstructs the global map.  

%\rev{1- primary list for shut down channel 2- secondary list based on overlap. it depends on the angle and location} 

\textbf{Local Map Reconstruction:}
With the base layer video received at each node, along with that node's own high quality 4K original video, each node can perform a quick attempt at the map reconstruction. First, images are compared pairwise using SIFT/ORB to determine feature matches. Some of these pairwise matches will be false, and will appear in some pairwise comparisons but not in others that show similar perspectives on the scene.
%
%To combat this, SfM includes a Random Sample Consensus~(RANSAC) process to remove false matches and curate the list of accepted feature matches. The SfM algorithm then uses these matches to simultaneous converge on relative camera position and the position of each feature match in 3D space. 
Because all nodes have some versions of the video, from different angles, the quality of reconstruction (measured by number of feature matches) 
should relate to two factors. Firstly, it depends on the amount of error-induced distortion in the base layer videos received from the other nodes. Secondly, it depends on the utility the locally stored 4K quality video on the reconstructing vehicle provides to the map reconstruction. Therefore, a vehicle that makes many feature matches in the intermediate local reconstruction attempt is a good candidate to share its recorded video at a higher quality in the next phase, because its video is a valuable part of the reconstruction and easy to match with the other videos.
%
%Maybe should name the final dense reconstructing vehicle. Final Reconstructor (FR)? Need better name.
%
%I think this is good but might not belong here. Challenges of getting valuable video underwater
The underwater environment poses additional challenges in recording good video for the purposes of map reconstruction. While 
%\blue{citation for FOV/underwater reconstruction paper} 
it can be shown that water itself is not a barrier to getting a good  reconstruction, there are serious problems with lighting, scattering, turbidity, and clarity when taking underwater video. %\textit{Even the problem of lighting an environment is difficult, with advanced post-processing techniques being required to remove headlights.} 

%\rev{M:what happens if mismatch happens?} \blue{SFM solves this, as we add more images it looks at agreement/disagreement about point placement in space} \rev{ok, then consider this situation in your algorithm and explain in the text.}

%\textbf{Channel Information Extraction:} 
%From the pilot information obtained along with the base layer video transmitted in the last step, each node is able to obtain an estimate of the channel to each other node in turn. This will be useful in both $(i)$~determining how many layers to transmit to that node in future enhancement layer transmission steps, as well as in $(ii)$~evaluating how good a candidate for sharing video each node is in the step that immediately follows. 

\textbf{Scoring and Sharing:} 
Using the optimizations described in the previous sections, each transmitting node decides on the set of nodes to shut down before broadcasts its higher quality layers, i.e., enhancement layers. Therefore, some nodes miss some portions of video from some other angles since they did not receive them. We form a Reconstruction Score~(RS) which is taken as a metric for how successful this vehicle would be at performing the later final reconstruction, as well as how valuable its local video is. 
This RS is shared in the following step to elect the Final Reconstructing Vehicle~(FRV).
%
%The number of points in the sparse reconstruction that come solely from matches within its own video sequence show that the locally-available video is generally suitable for reconstruction, in that it must have clearly discernible features of the object visible and consistent between frames. However, the number of outside matches (matches made between the locally available video and ones downloaded from other nodes) indicates that the video exists within the same context as other videos and can contribute to the assembly of the full model if shared. For that reason, points that are included in frames of video from other vehicles are weighted higher in the SRS. The vehicle with the highest SRS will be elected as the FRV.
%
Each node will share its RS to the group, such that at the end of this step all vehicles should have a list of each other vehicle's RS. As the process continues, nodes will become more aware of their position relative to other nodes. Since the RS is a very small amount of data, each vehicle can also share in the packet a map of camera positions (past vehicle positions) it has matched with. An average of these maps can be used to inform the vehicle's navigation in the time before the final reconstruction can be performed.
%
%In this work, the RS is used as the sole criterion for selecting the FRV. In future, other useful factors include the average channel capacity to each vehicle and its energy level, because dense reconstruction is an energy intensive process and it would be advantageous to select a vehicle with enough battery life left to support the process.
%
%\rev{M: do they need to share their location as well? or this comes from sparse reconst.?}
%\blue{If the reconstruction works, their relative positions can be found very well since SFM places the cameras to perform the reconstruction. So we don't need external positioning unless we can expect the sparse reconstructions are not functional at early stages.} \rev{ok, please explain this in the text, please.}

%I'm not sure if this step is important anymore? Should we incorporate CSI in the decision at this step?
%
%

%%%%%%%%%%%%%%%%%%%%%%%%%%%%%%%%%%%%
\begin{algorithm}[!t]
\caption{SVC-based Map Reconstruction.} \label{algo:proc}
%\todo{doublecheck for consistency}
%\todo{add comments after each code line}
\begin{algorithmic}[1] 
%\small
 	%\STATE{//algorithm runs on vehicle $v$}
 	\WHILE{reconstruction is NOT satisfactory}
 	\STATE{Layers $=$ ScalableVideoCoder(localVideo)}
 	\STATE{EstablishMACchedule()}
 	\STATE{$\tau_b$ $\leftarrow$ allotted time for base layer sharing}
 	\WHILE{$t<\tau_b$}
 	%\IF{$t\in \rm{slot_v}$}
    \STATE{Share(Layers.LayerIndex(0))\hspace{0.99cm}{\% broadcasting}}
    %\ELSE
    \STATE{Receive(ExternalVideo)}
    %\ENDIF
    \ENDWHILE
    %\STATE{Decide on $\alpha_i$ and $R_i$, for $i=1,...,L+1$, based on~\eqref{opt} }
    %\STATE{Transmit(baseLayer); $s \leftarrow 1 $ \quad \quad \%  $s$ is the number of trials}
    \STATE{receivedframes $ \leftarrow $ extractframes(receivedVideos)}
    \STATE{SIFT/ORBmatch(receivedframes)}
    %\STATE{ORBmatch(receivedframes)}
    \STATE{Reconstruct(matchedframes)}
    \STATE{RS $\leftarrow$ score(reconstruct)}
    \STATE{random\_broadcast\_max(RS)}
    \STATE{$\tau_l$ $\leftarrow$ allotted time for enhancement layer sharing}
 	\WHILE{$t<\tau_l$}
    \IF{$v$ is not FRV}
    %\STATE{find maximum no. of layers}
    \STATE{Shut down the vehicles with the weakest channel}
    \STATE{Share(Layers.LayerIndex($L$)) \hspace{0.8cm}{\% multicasting}}
    \STATE{Receive(ExternalVideo)}
    \ELSE
        \STATE{Reconstruct(Dataset)}
       % \STATE{$v$ leads mission} %Not sure how to statej this in Pseudocode
    \ENDIF
    \ENDWHILE
    \ENDWHILE
    \end{algorithmic}
 \end{algorithm}
    
%    \IF{feedback is resend request}
 %   	\STATE{transmit(feedback.requestedLayer)}
 %       \STATE{channelState.update(BAD)}
 %        \STATE{$s \leftarrow s+1$}
  % \ELSE
 %   \STATE{transmit(Layers.nextLayer)}
 %   \STATE{$s \leftarrow s+1$}
 %   \STATE{channelState.update(GOOD)}
%    \ENDIF
  %  \IF{channelState.rollingAverage $>$ threshold}
 %\STATE{transmitter.switchTo('Multiplexing')}
 %\ENDIF
 %Get data and make packet(); store packet()
% \STATE{Step~II:}
% \FOR{antennae set $(k,l)$, serial/parallel pipes}
% \STATE{Segmentation $(b_n)+(a_n)$}
% \STATE {Signal mapping $(b_n)$, QPSK in our case}
% \STATE {Space mapping $(a_n)$ according to Fig.~\ref{fig:block_trans_1}}
% \STATE{Call IFFT ()}
% \STATE{Call CP ()}
% \STATE{Form transmission Matrix (N, subcarriers)}
% \STATE {Transmission Matrix(Subcarrier ON/OFF info)}
% \STATE{Beamforming $F(\theta_{k_{V-B}},\phi_{k_{V-B}})$}
% \STATE{ Transmitted signal $\textbf{v}\big(t,\theta_{k_{V-B}},\phi_{k_{V-B}})$}
% \ENDFOR
% \ENDIF

%\textbf{Consensus Algorithm on the Scores and Election of Reconstructing Vehicle:}
%Consensus algorithm deals with converging an agreement between the connected nodes in an error-prone communication channel.
\textbf{Consensus Algorithm on the Scores:}
To select the vehicle with the highest score for the final reconstruction, vehicles form the communication primitive to their neighboring vehicles. In particular, consensus is an iterative process where the nodes communicate with their neighbors to exchange their scores for a fixed number of iterations or until convergence~\cite{pandey2018robust}. As the output of this process, we select the best vehicle for final reconstruction.
%and the amount of required data 
Asynchronous broadcasting-based consensus method proposed in~\cite{pandey2018robust} is to achieve the average value of the initial measurements. However, we wish to sort the scores to find the maximum in each iteration of the process. Each node $v$ broadcasts its own score to its $\mathcal{N}_v$ neighboring nodes within its communication range~\cite{iutzeler2012analysis}. 
%The network is assumed asynchronous, i.e., there is no common clock available and so each node has its own clock and can initiate the broadcasting at its clock tick~\cite{iutzeler2012analysis}. 
%Assume each node has a clock which ticks independently according to a Poisson process
%~\cite{aysal2009broadcast}.
The neighbors, such as $w$, which received the data, update their data according to $y_w{(t_c+1)}= \rm{max} \big(y_v{(t_c)},y_w{(t_c)}\big), \forall w \in \mathcal{N}_v$, where $N_v$ stands for the neighborhood of transmitting node $v$. The remaining nodes in the network update their values as $y_w{(t_c+1)}=y_w{(t_c)}, \forall w \notin \mathcal{N}_v$. This algorithm keeps the maximum value and so does not show an undesirable behavior in terms of convergence. 
%
%The neighbors, which received the data, update their data according to the weighted average of their current data as, $y_v{(t_c+1)}=\gamma y_v{(t_c)}+(1-\gamma) y_i{(t_c)}, \forall v \in \mathcal{N}_i$, where $\gamma \in (0,1)$ stands for the mixing parameter, which gives the weight assigned to data values from each node, and $\mathcal{N}_i$ denotes the neighboring nodes of the $i^{th}$ node. The remaining nodes in the network update their values as $y_v{(t_c+1)}=y_v{(t_c)}, \forall v \notin \mathcal{N}_i$. Each node achieves perfect consensus averaging as $t_c \rightarrow \infty$, where $t_c$ is the number of consensus iterations, and obtains $Y_{i,T}{{(\infty)}^{\top}}=\frac{1}{N} \sum_{j=1}^{N} y_j{(0)}$~\cite{pandey2018robust}.
%
%
%\textbf{Election of Final Reconstructing Vehicle:}
After consensus, each vehicle should know the maximum RS among them and the vehicle that has it. The vehicle who has the highest score will transmit a final packet indicating its RS and intent to become the FRV. If there is no reply within the time limit, it is the FRV and the SVC enhancement layer sharing will commence.
Algorithm~\ref{algo:proc} represents the solution in a sequential procedure for a specific coded video while the encoding and reconstruction is performed through the mentioned steps. Vehicles share their encoded base-layer and enhancement layers videos (after shutting down the vehicles with a low quality channel). After local reconstruction, matching and ranking the scores, the node with the highest score will be elected to perform the final reconstruction. 
\begin{figure}[t!]
\centering
\begin{tabular}{cc}
\includegraphics [width=0.23\textwidth]{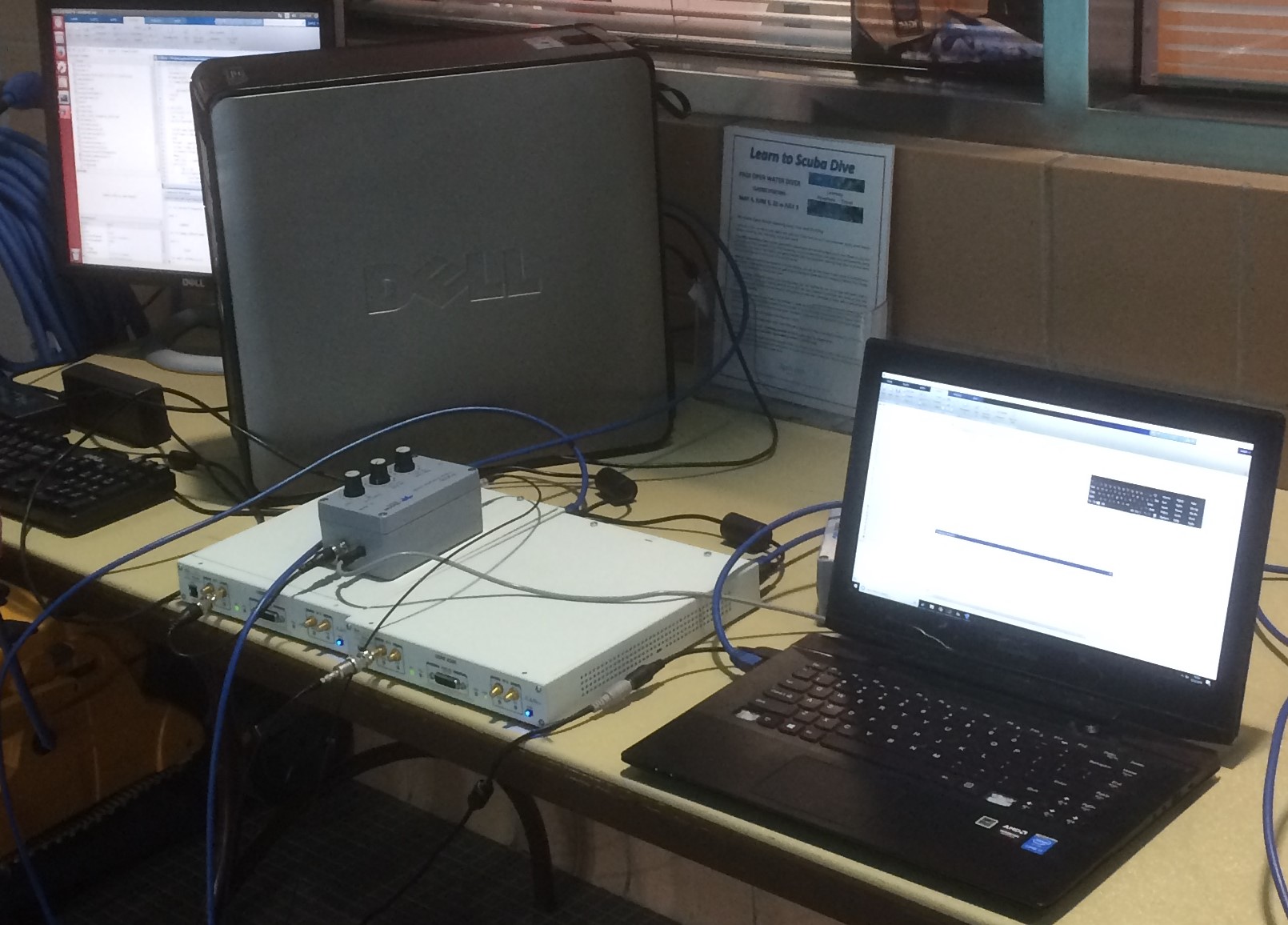}
\includegraphics [angle=90,height=3cm, width=0.23\textwidth]{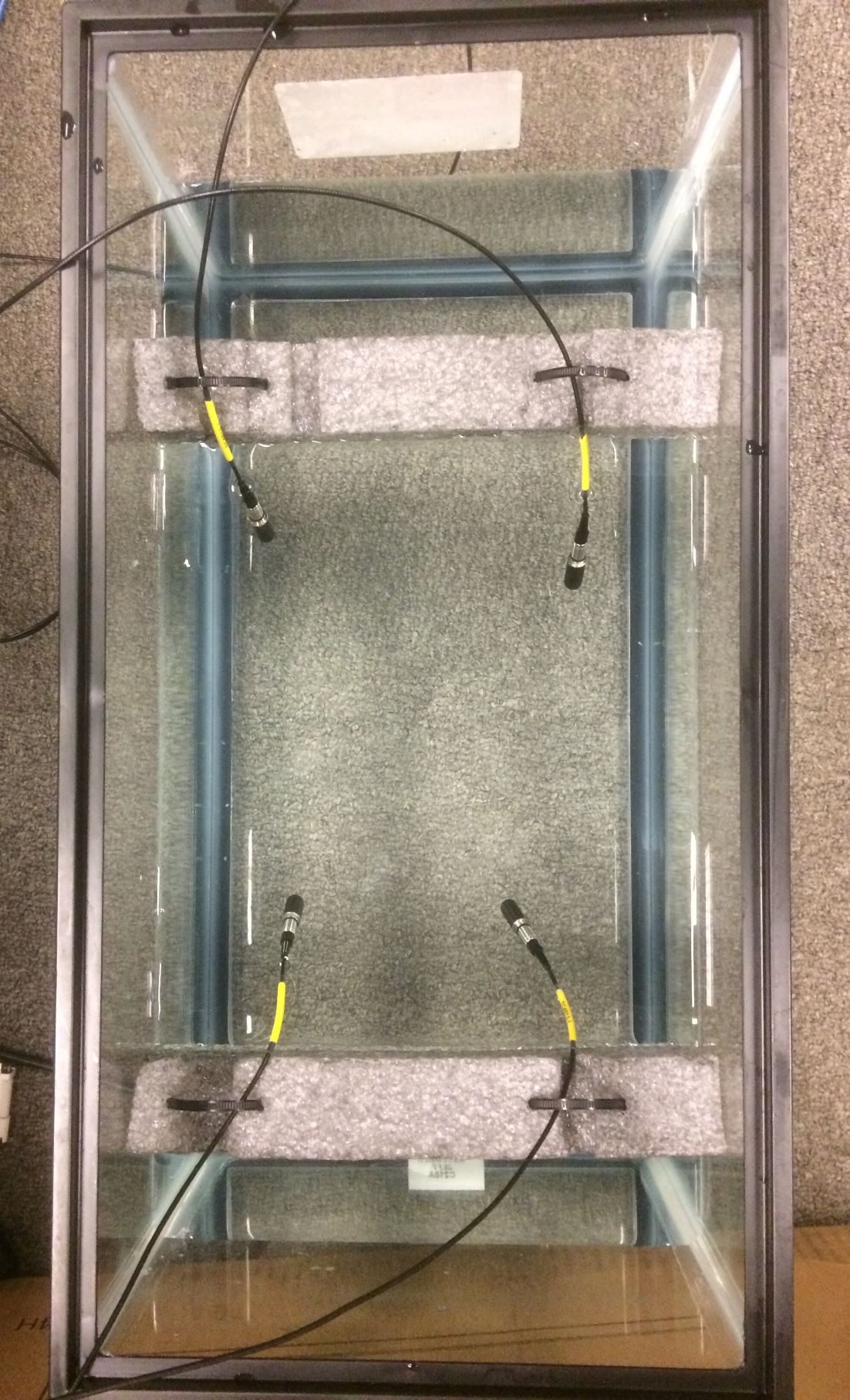} \\
\hspace{0.5cm} (a)   \hspace{3.5cm} (b) 
\end{tabular}
\caption{(a) Software-defined acoustic testbed; (b) Water tank with TC4013 Teledyne transducers.}\label{fig:testbed}
%\vspace{-2mm}
\end{figure}
%%%%%%%%%%%%%%%%%%%%%%

%%%%%%%%%%%%%%%%%%%%%%%%%
\begin{figure}[t!]
\centering
\begin{tabular}{cc}
\hspace{-5mm}
\includegraphics [width=0.26\textwidth]{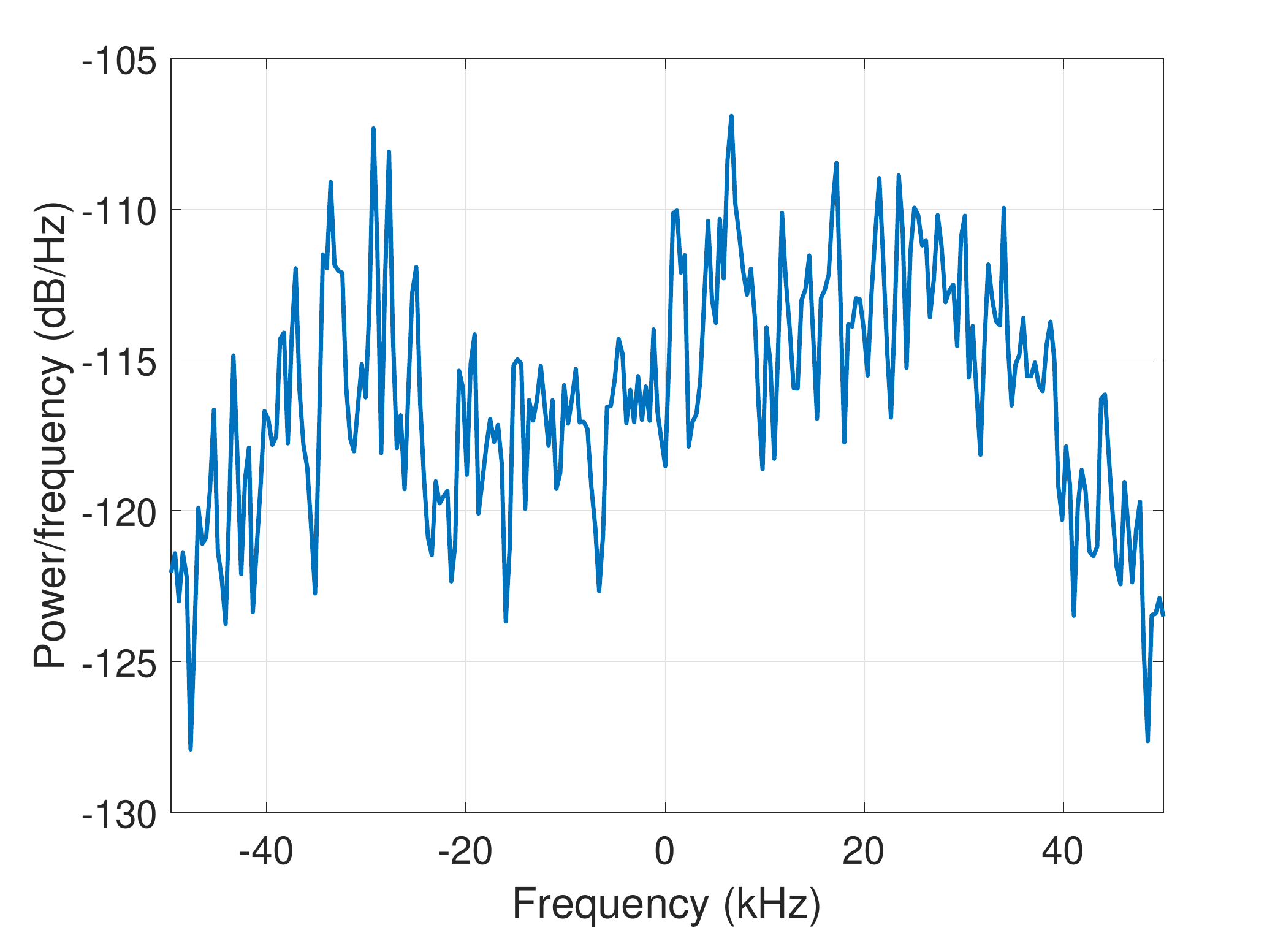} \hspace{-0.65cm}
\includegraphics [width=0.27\textwidth]{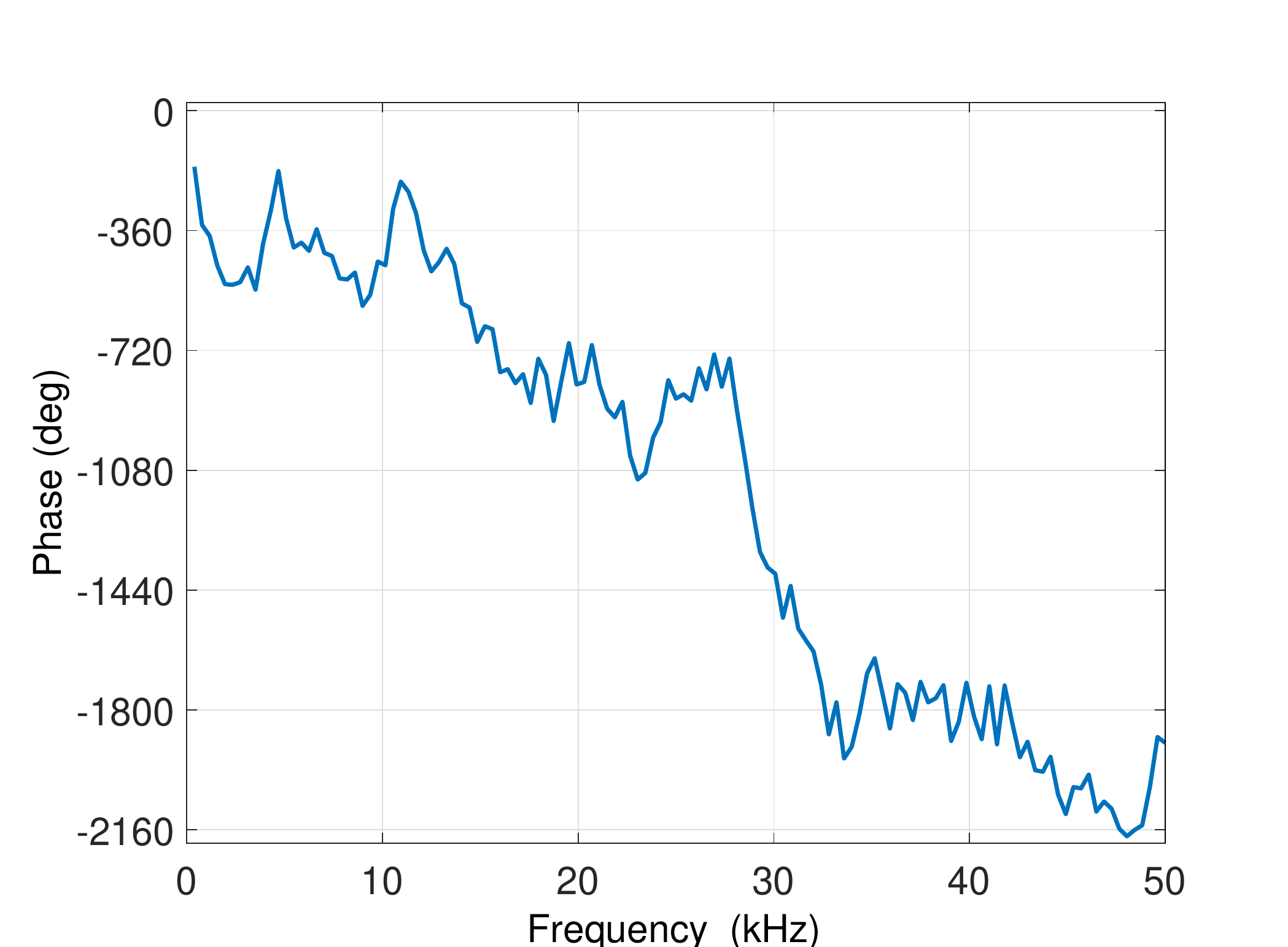}\\
\hspace{0.5cm} (a)   \hspace{3.5cm} (b) 
\end{tabular}
\caption{Experimental channel response in the water tank shows (a)~Power spectrum; (b)~Phase.}\label{fig:channelresponse}
%\vspace{-2mm}
\end{figure}
%%%%%%%%%%%%%%%%%%%%%%

\begin{figure}[!t]
\centering
%\setlength{\intextsep}{-1pt}
%\par\vspace{\intextsep}
\begin{tabular}{ccc}
\hspace{-0.29in}
\includegraphics[width=0.36\columnwidth]{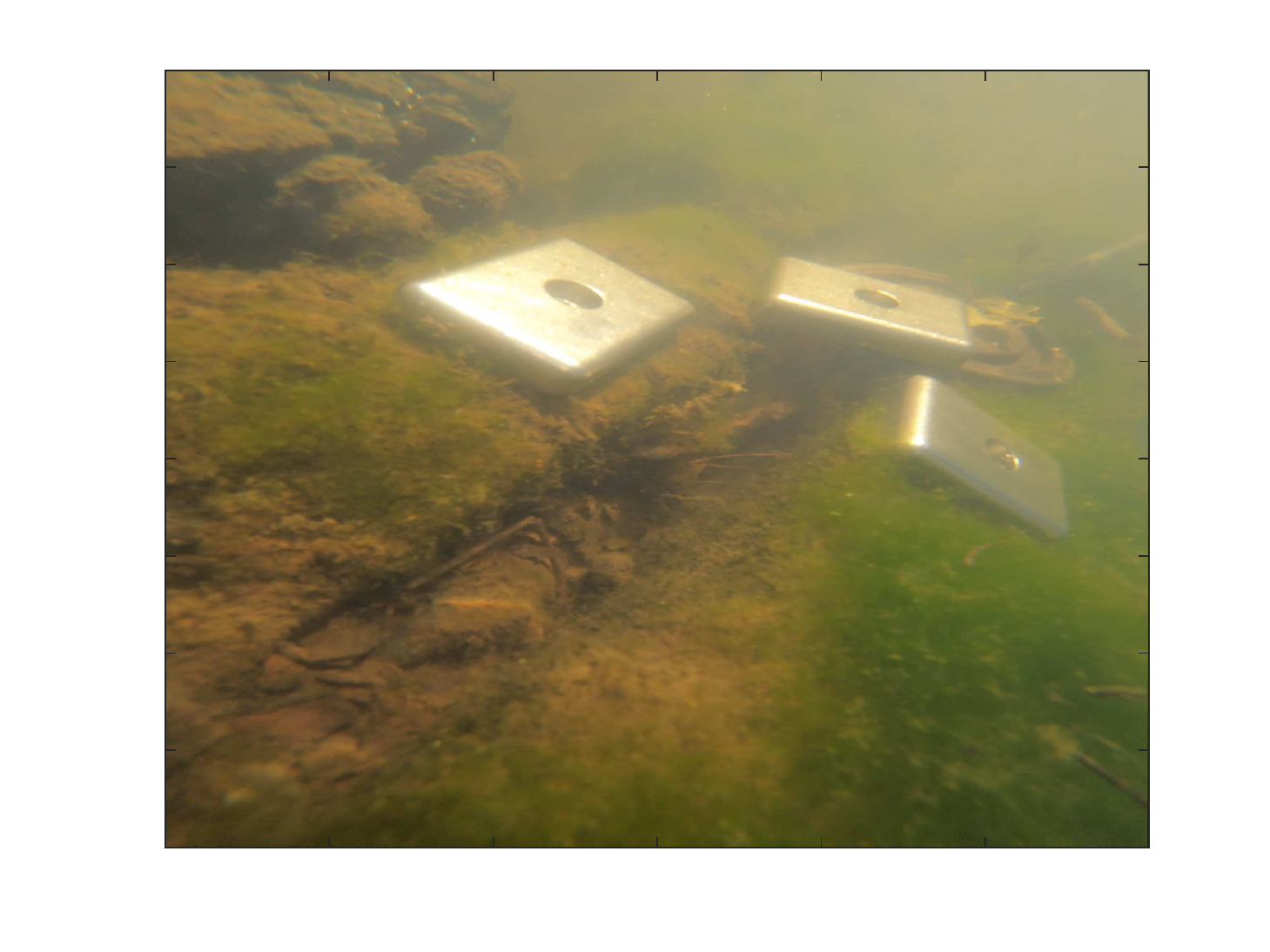} &
\hspace{-6mm}
\includegraphics[width=0.36\columnwidth]{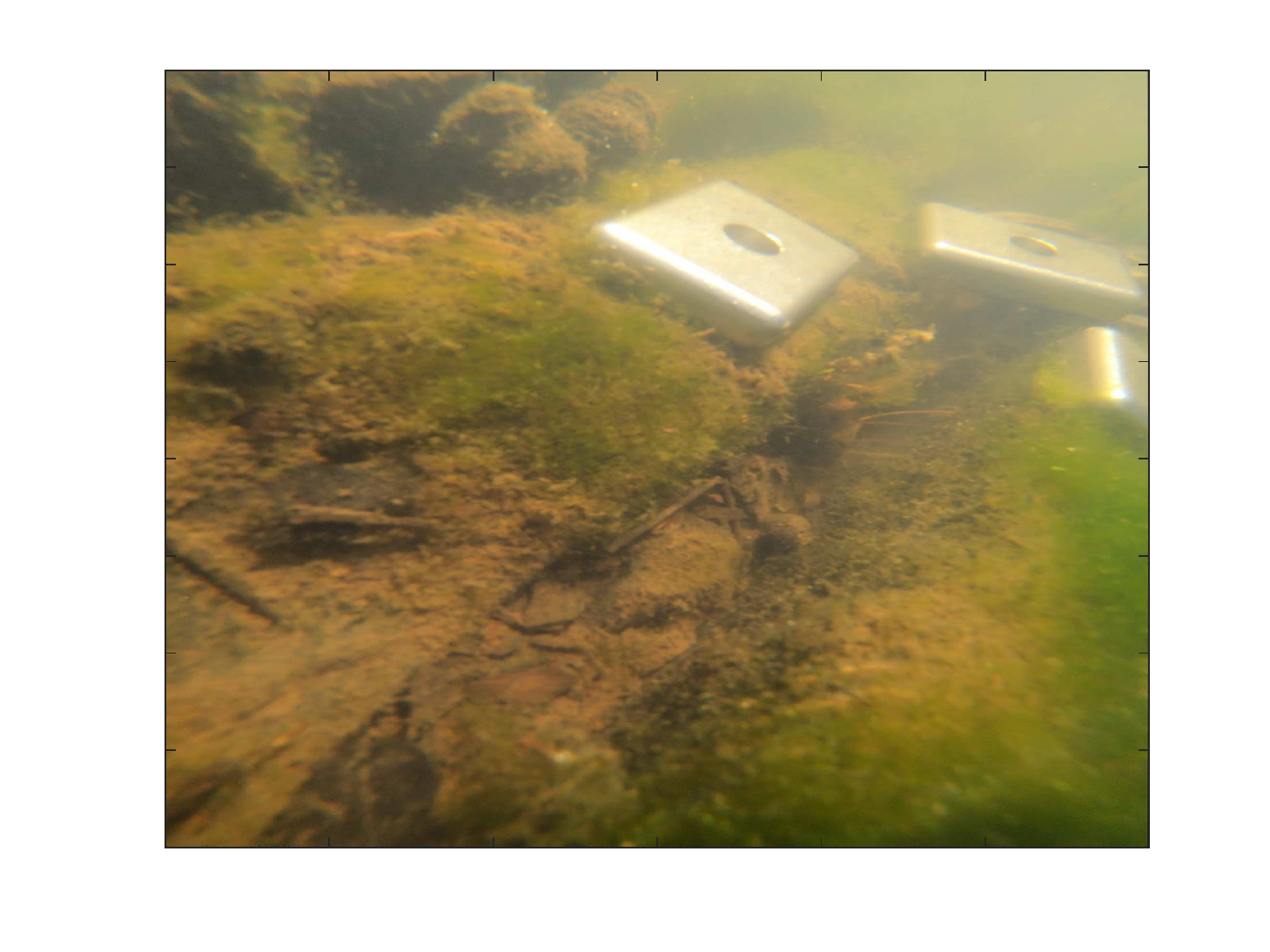} &
\hspace{-6mm}
\includegraphics[width=0.36\columnwidth]{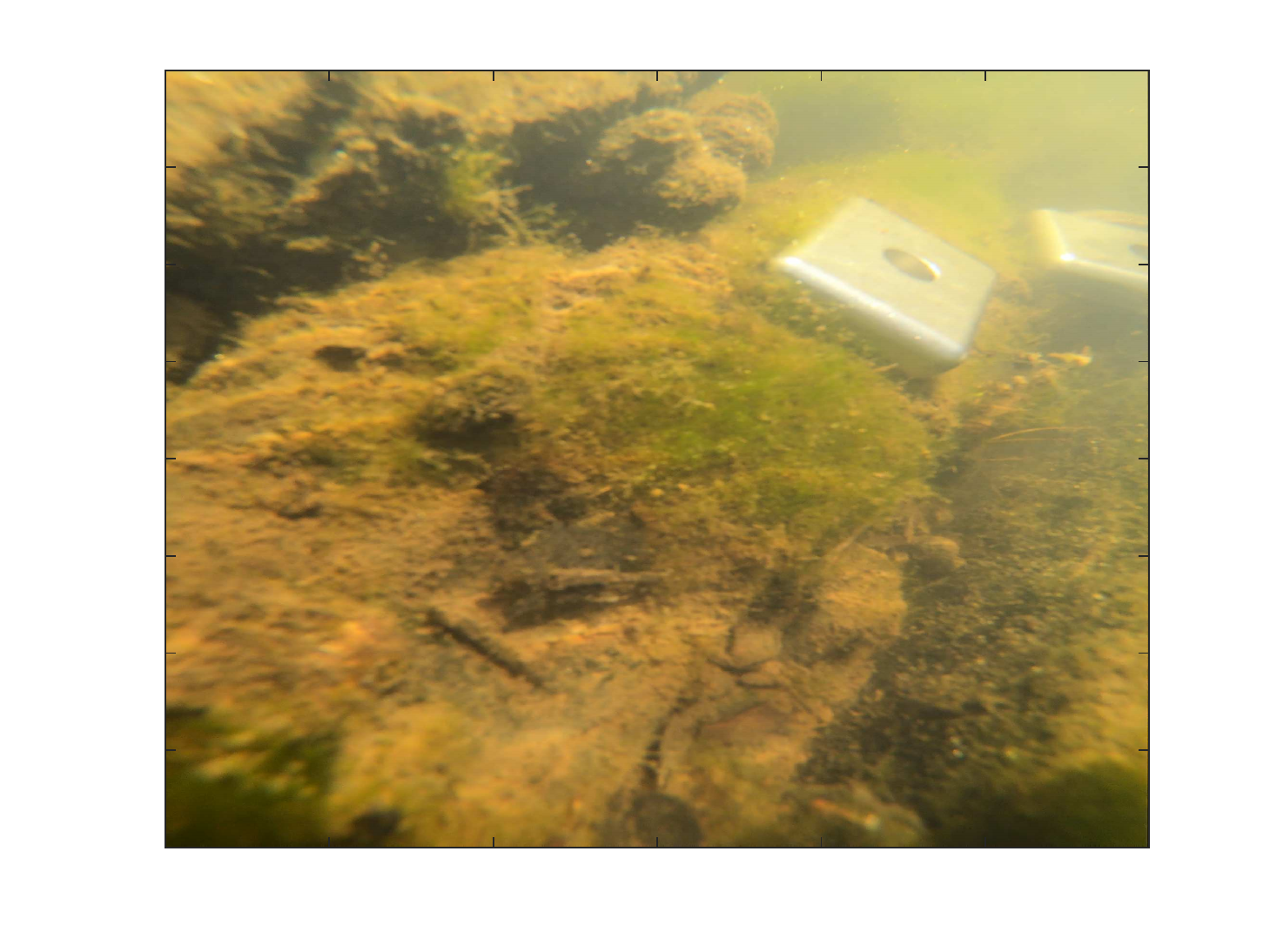} \\
%\vspace{-0.35cm}\\
%\vspace{-6mm}
%\vspace{-0.2 cm}
%%%%%\small (a) & \small (b)  & \small (c)
\small (a) & \small (b) &  \small (c)
\end{tabular}
%\vspace{5mm} 
%\hspace{0.25in}
\begin{tabular}{ccc}
\hspace{-0.29in}
\includegraphics[width=0.39\columnwidth]{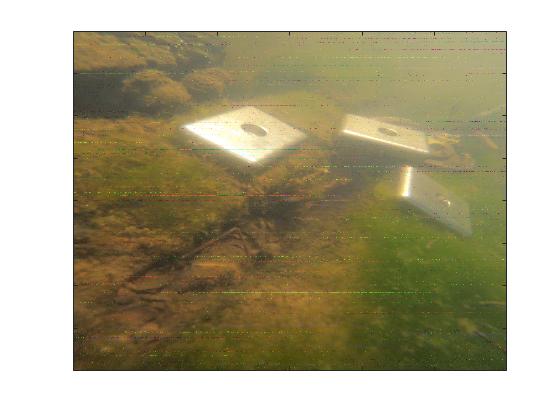} &
\hspace{-9mm}
\includegraphics[width=0.38\columnwidth]{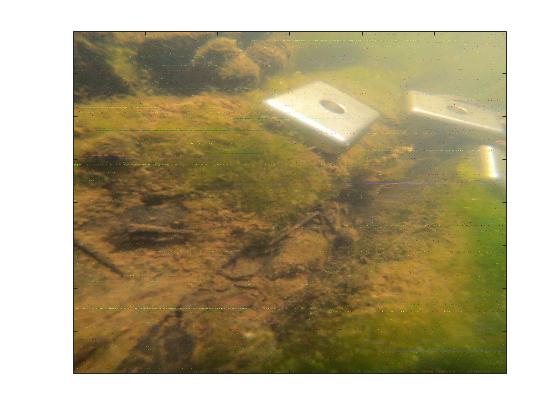} &
\hspace{-9mm}
\includegraphics[width=0.38\columnwidth]{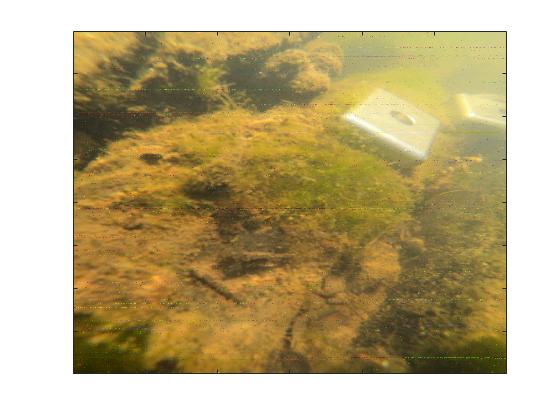} \\
 \small (d) & \small (e) &  \small (f)
\end{tabular}
\begin{tabular}{ccc}
\hspace{-0.29in}
\includegraphics[width=0.39\columnwidth]{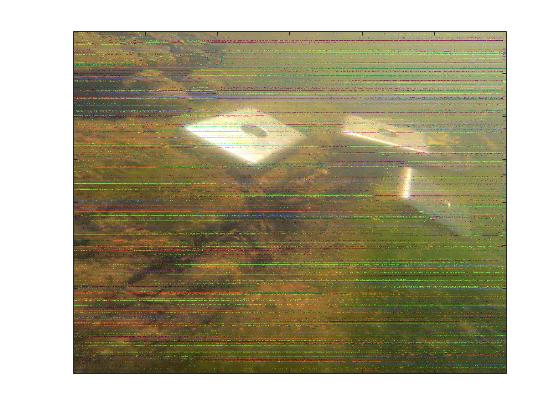} &
\hspace{-9mm}
\includegraphics[width=0.38\columnwidth]{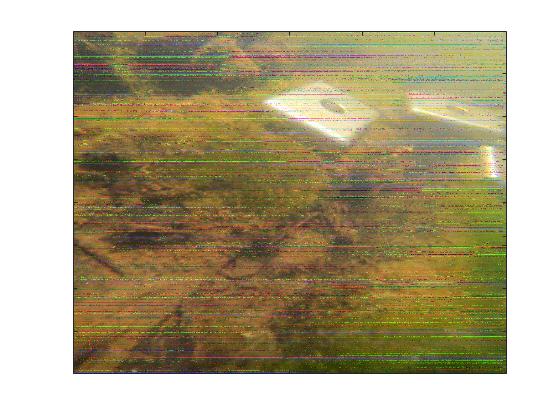} &
\hspace{-9mm}
\includegraphics[width=0.38\columnwidth]{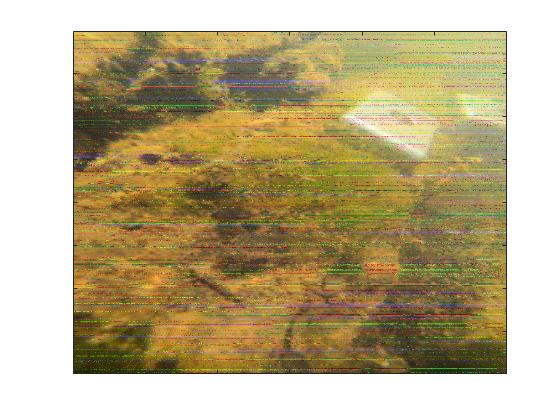} \\
 \small (g) & \small (h) &  \small (i)
\end{tabular}
\caption{(a)-(c)~Frames from original video; (d)-(f)~Frames of video received/reconstructed in a vehicle with a good channel; (g)-(i)~Frames of video received/reconstructed at a vehicle with an average to low channel quality.\label{fig:Eval2}}
\vspace{-2mm}
\end{figure}

\begin{figure*}[!t]
\centering
\setlength{\intextsep}{-1pt}
\par\vspace{\intextsep}
\begin{tabular}{ccccc}
\hspace{-0.24in}
\includegraphics[width=0.19\textwidth]{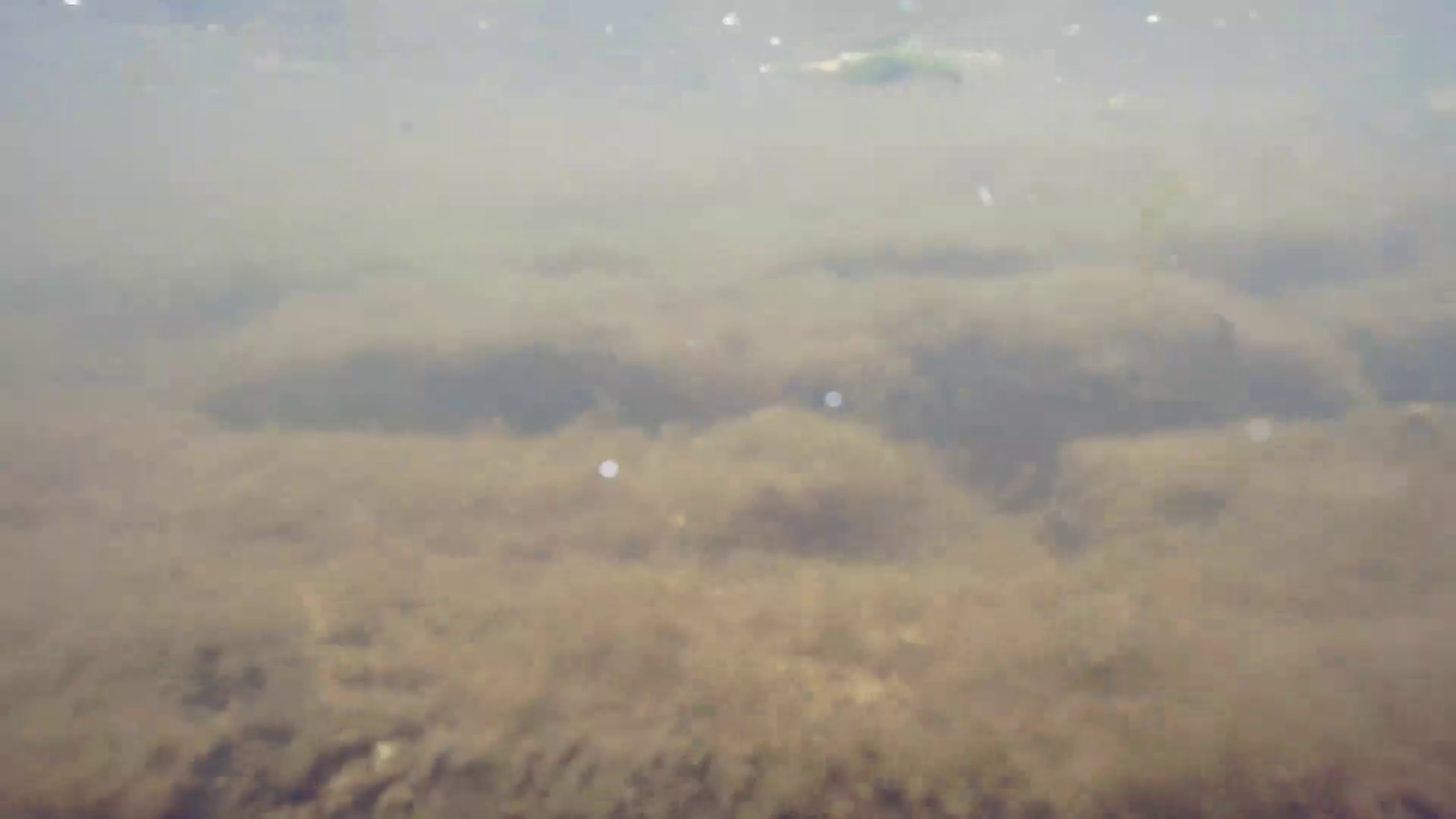} &
\hspace{-3mm}
\includegraphics[width=0.19\textwidth]{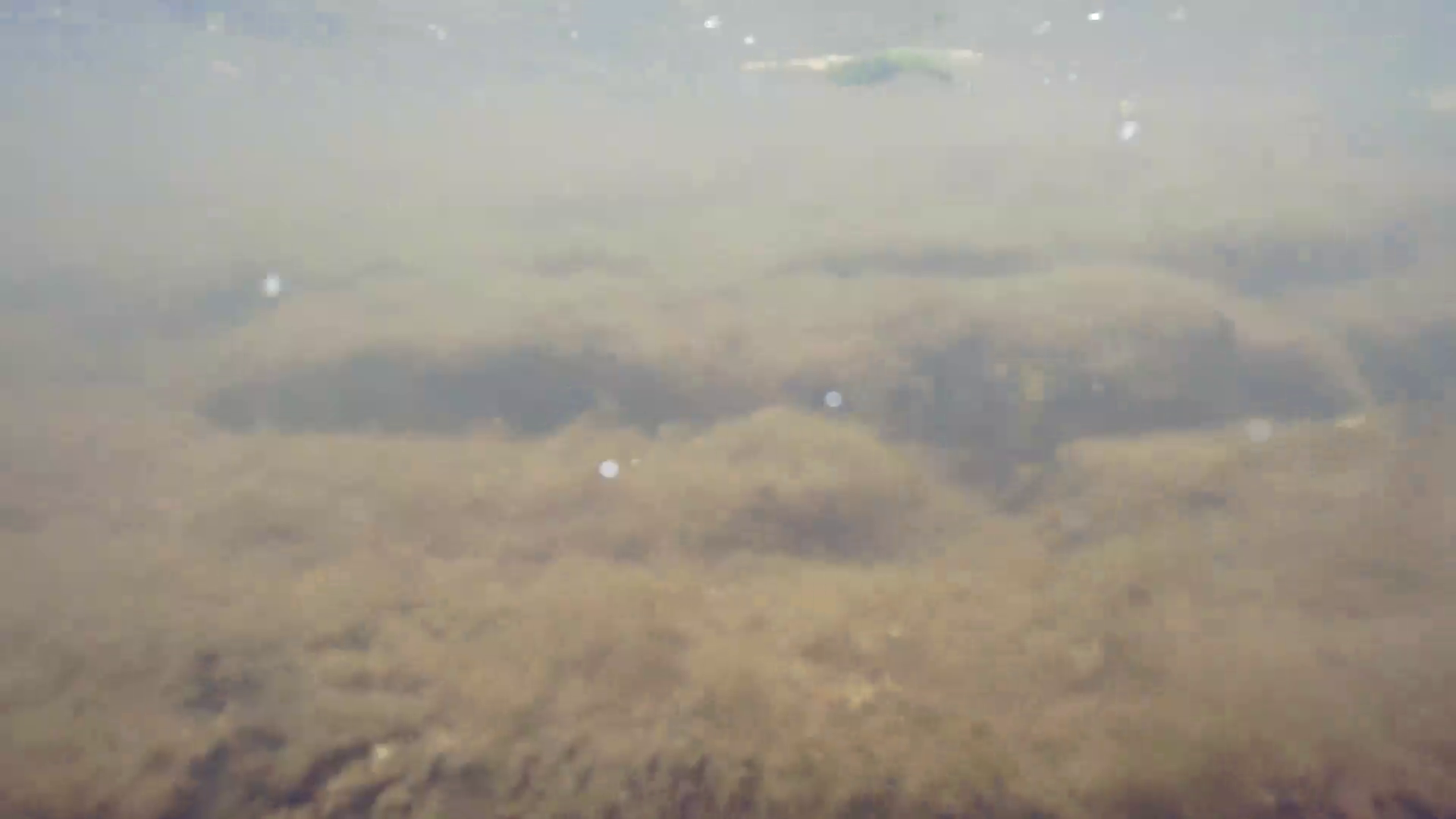} &
\hspace{-3mm}
\includegraphics[width=0.19\textwidth]{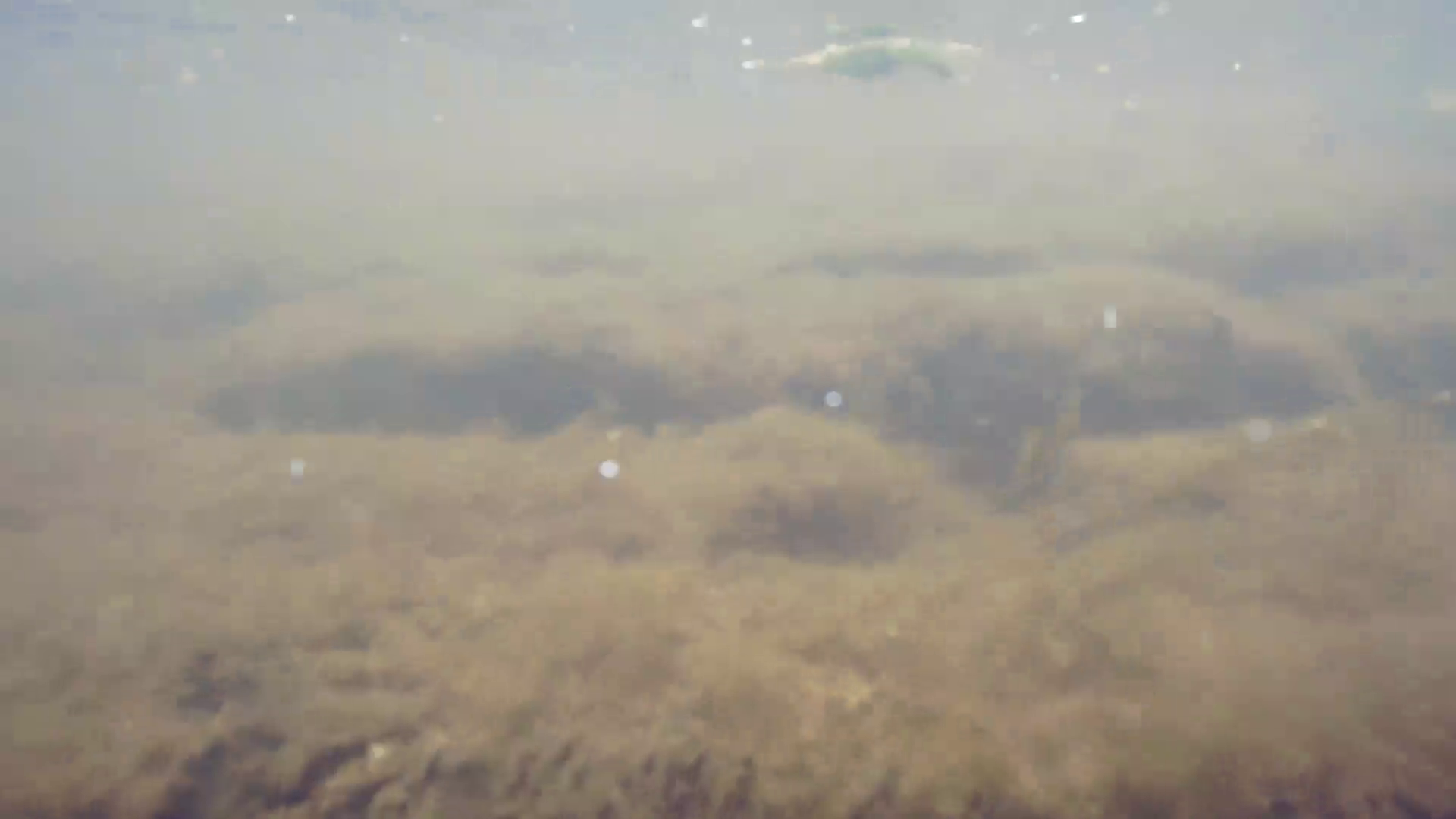} &
\hspace{-3mm}
\includegraphics[width=0.19\textwidth]{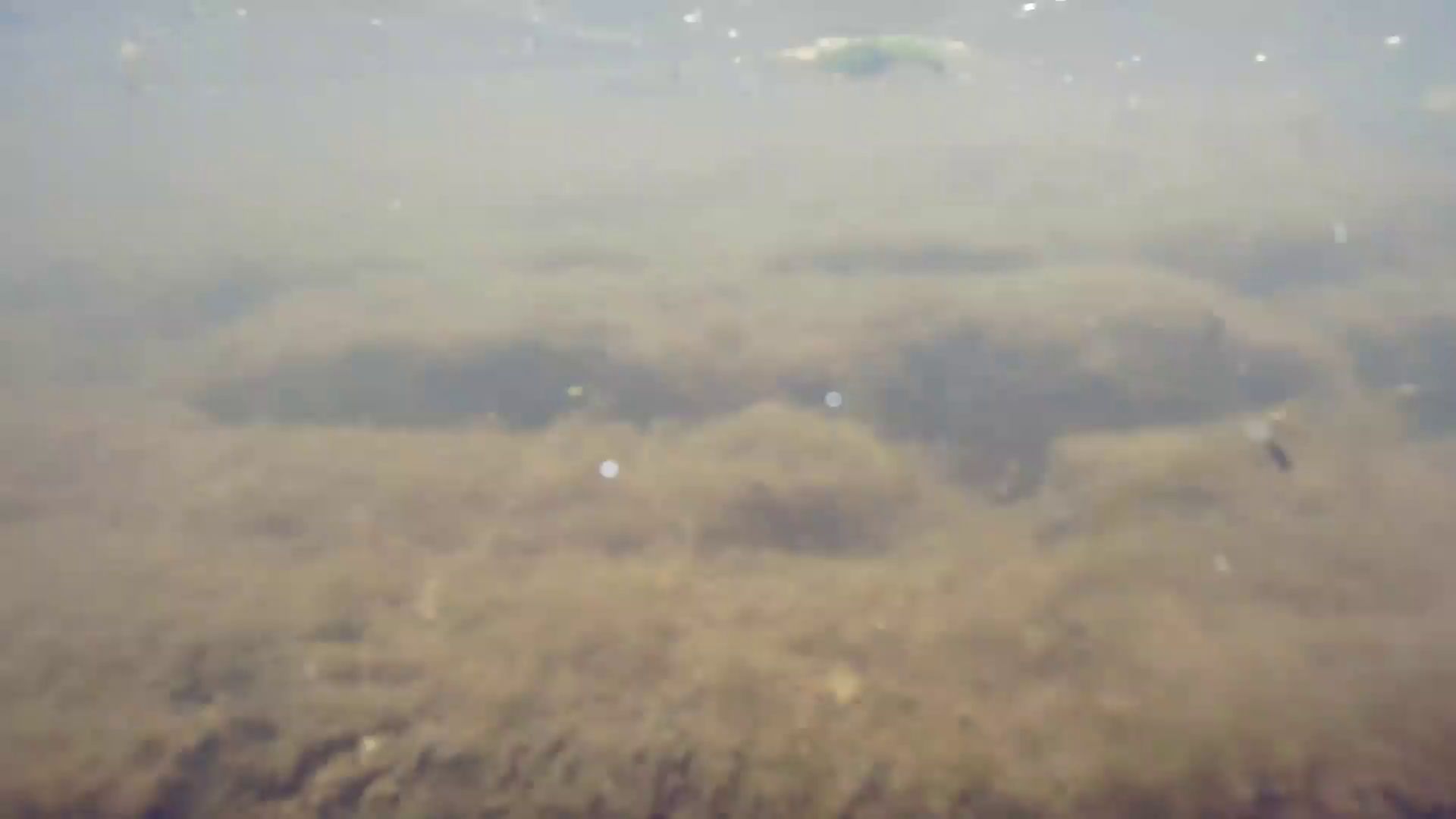} &
\hspace{-3mm}
\includegraphics[width=0.19\textwidth]{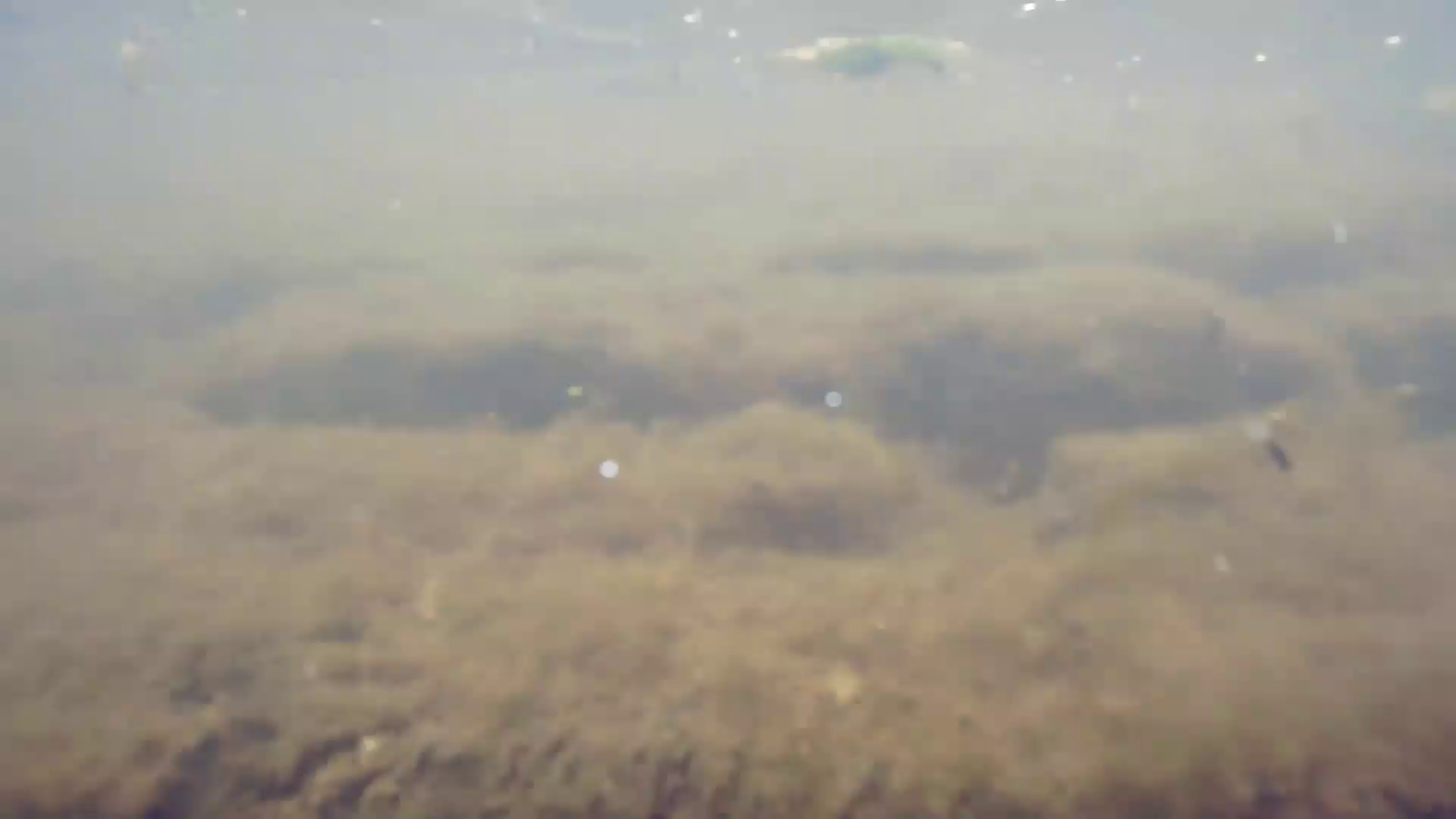}\\
\small (a) & \small (b) &  \small (c) &  \small (d) &  \small (e)
\end{tabular}
\begin{tabular}{ccccc}
\hspace{-0.24in}
\includegraphics[width=0.19\textwidth]{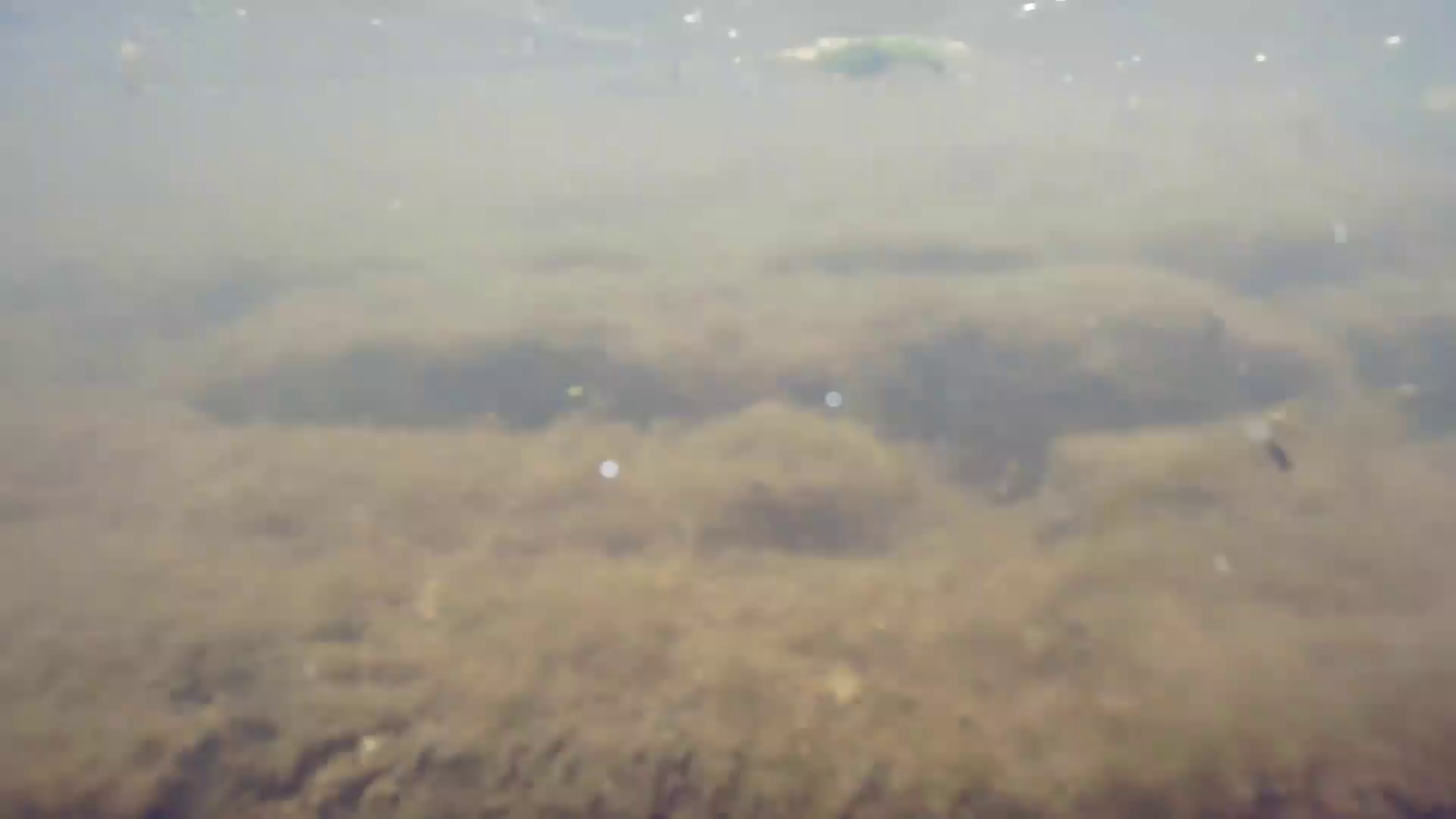} &
\hspace{-3mm}
\includegraphics[width=0.19\textwidth]{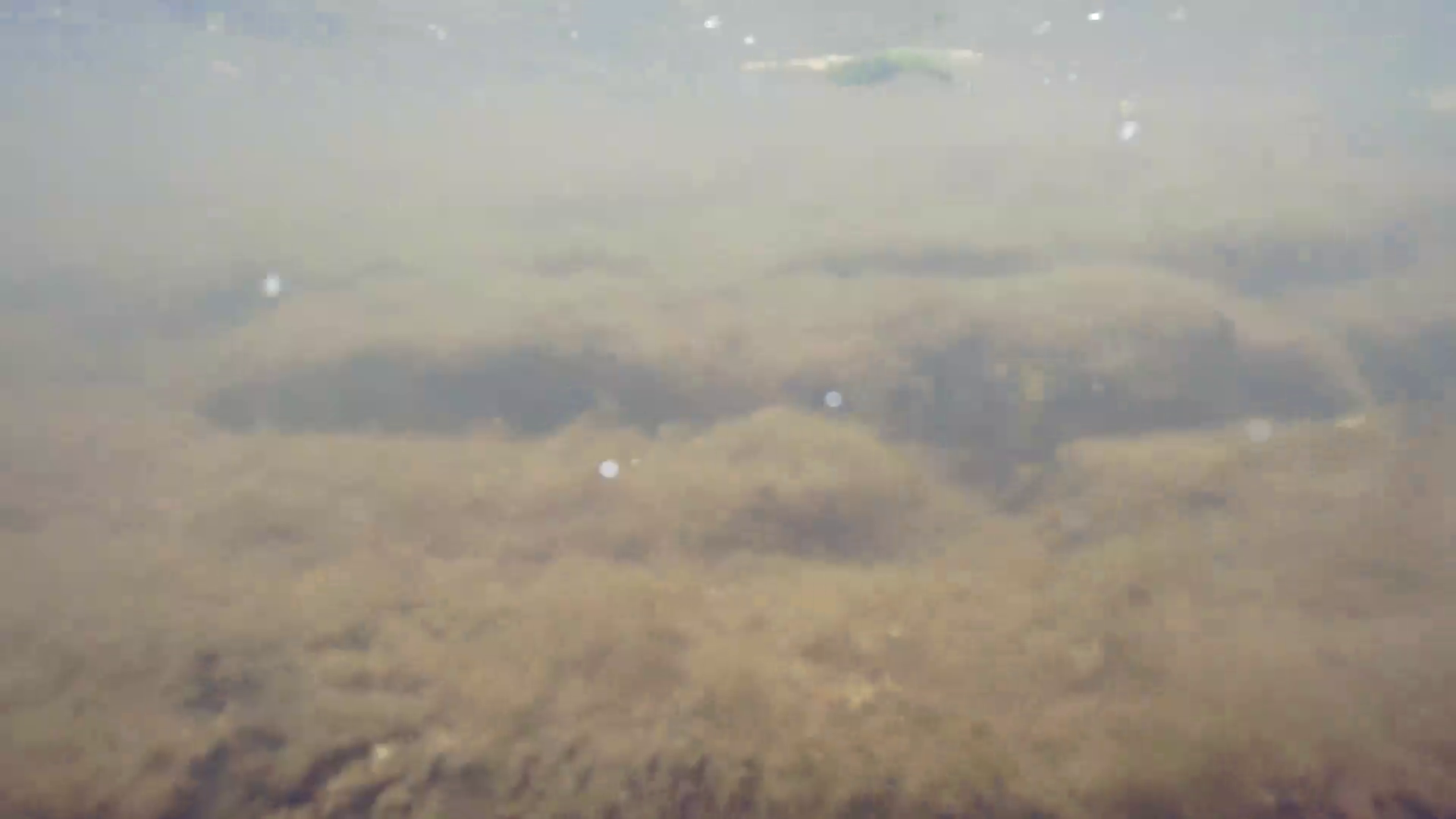} &
\hspace{-3mm}
\includegraphics[width=0.19\textwidth]{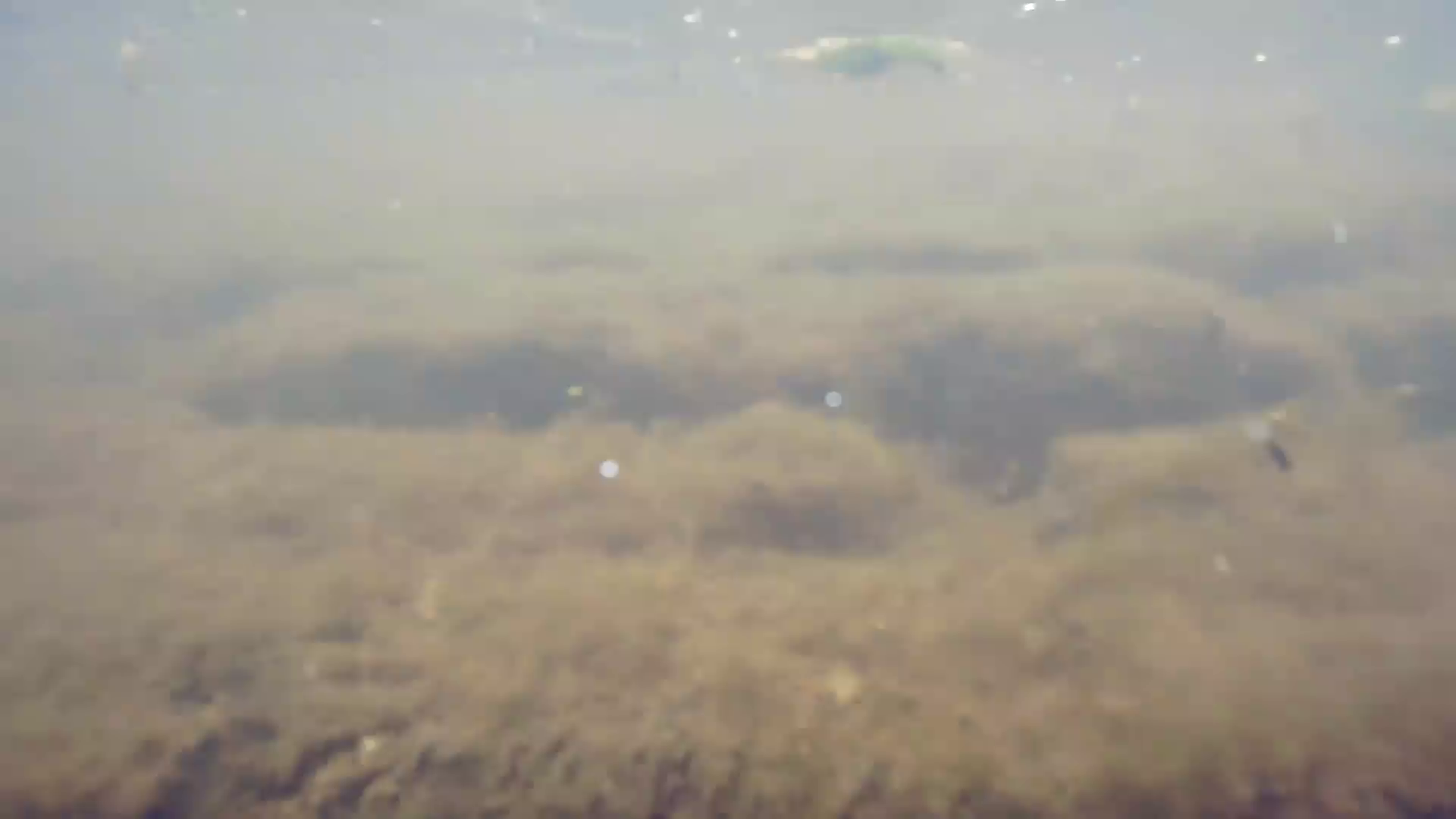} &
\hspace{-3mm}
\includegraphics[width=0.19\textwidth]{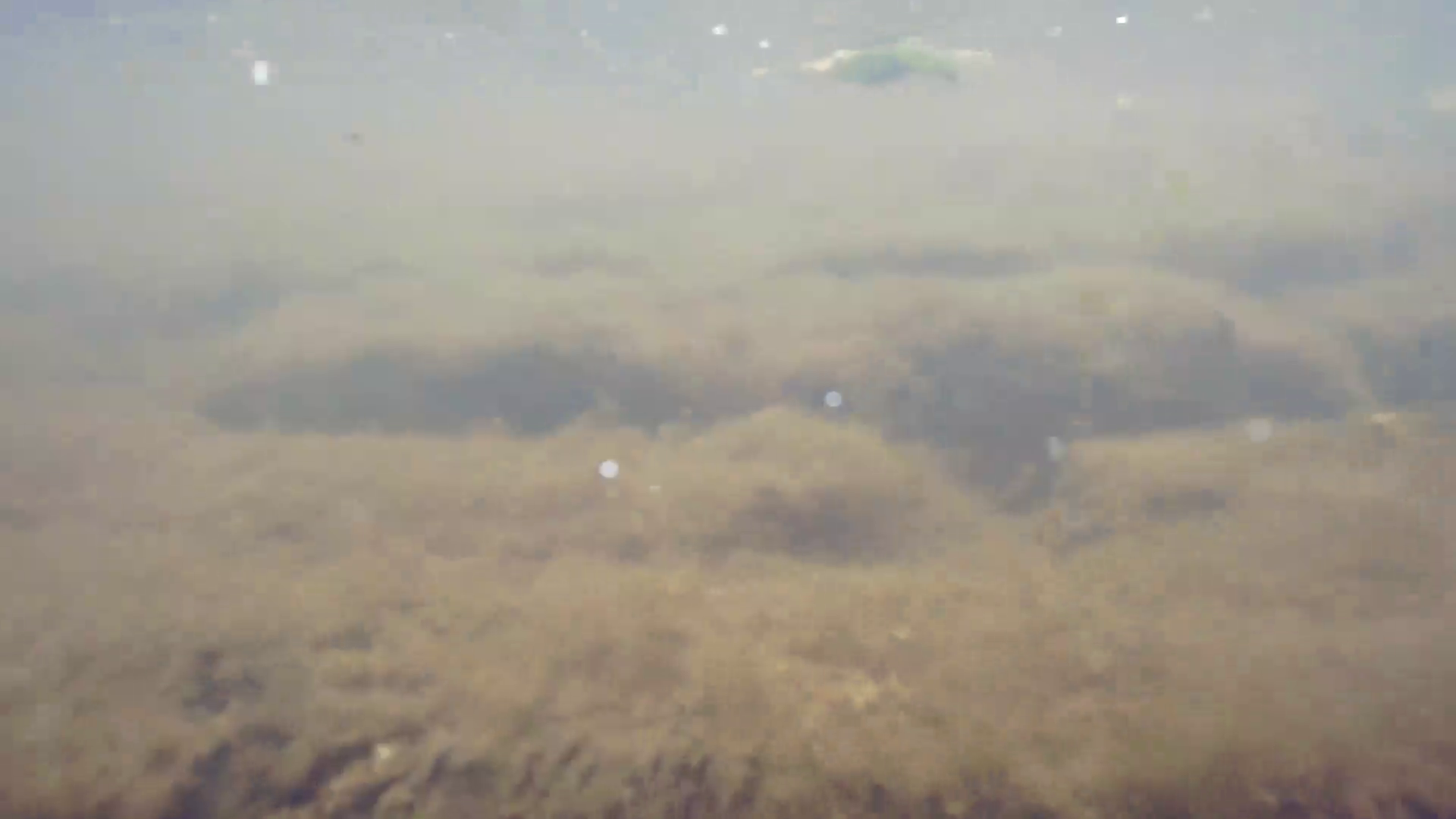}&
\hspace{-3mm}
\includegraphics[width=0.19\textwidth]{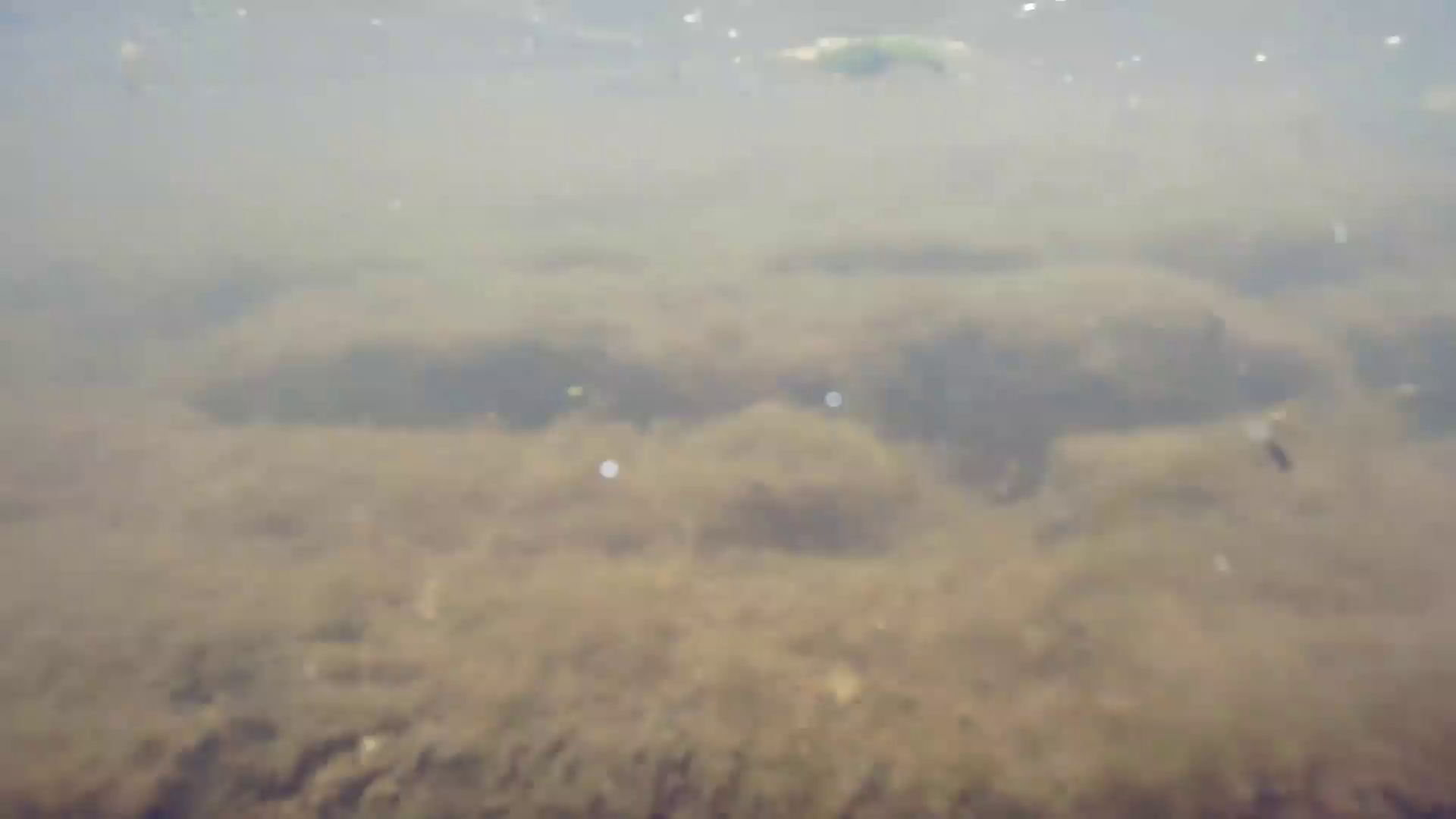}\\
 \small (f) & \small (g) &  \small (h) &  \small (i) &  \small (j)
\end{tabular}

\caption{SVC layers for a selected frame; (a)~base layer of original video; (b)-(e)~base layer and $1-4$ enhancement layers of original video; (f)~base layer of received video; (g)-(j)~base layer and $1-4$ enhancement layers of received video. \label{fig:svc_layers}} 
%\vspace{-1mm}
\end{figure*}

\section{Performance Evaluation}\label{sec:Eval}
In this section, we present the experimental and simulation results to evaluate the proposed algorithm.

%\todo{if we really have both experiments and simulations, then create two separate subsections, possibly to mirror the two in Sect. 3 (on video and comms, respectively)} 

\textbf{Testbed Setup:}
We evaluate the proposed approach by conducting preliminary field experiments.
A video feed, captured by our underwater vehicles in the Raritan river, New Jersey, is passed to the SDAR and an acoustic transducer in a water tank. A high-performance and scalable platform with a programmable Kintex-7 FPGA, called X-300 designed by Ettus Research Group~\cite{USRP}, is exploited as SDAR in this research, as the testbed shown in Fig.~\ref{fig:testbed}(a). It contains a mainboard to provide basic functionalities of the modem, while the daughter-boards take care of up/down conversions and of the other required bandpass signal processing procedures. Teledyne Marine TC4013 transducers~\cite{Tc4013} with a frequency range of $170~\rm{kHz}$ are used in the proposed testbed, shown in Fig.~\ref{fig:testbed}(b). Fig.~\ref{fig:channelresponse} represents the channel response experienced in this testbed, containing the power spectrum of the channel in~\ref{fig:channelresponse}(a) and its phase in~\ref{fig:channelresponse}(b). 
%
%\textbf{Settings:}
The video was collected from the bottom of the Raritan river, New Jersey, using multiple cameras. 
%The communications channels were simulated in MATLAB, assuming there are different qualities of channel between each vehicle and other neighboring vehicles, so that the frames of video are received with various qualities. 
We use the Joint Scalable Video Mode~(JSVM) software as the reference package for implementing SVC. Using the FixedQPEncoder program, test videos were down-sampled and then encoded into multiple layers of different qualities. Each layer has a target fixed bit rate, and the Quantization Parameter~(QP) is varied in order to optimize the Peak Signal-to-Noise Ratio~(PSNR) metric while staying under the target bitrate. %Additionally, 2 different temporal layers can be extracted for each quality layer, giving $6$ total layers. 
%Open SVC Decoder is used for decoding. %due to its implementation of error concealment and its integration with Mplayer for video streaming. Table~\ref{tab:SVCQoS} represents the quality of each encoding mode based on the number of layers in the encoding procedure.

%\textbf{Setup and Simulation Parameters:}

\begin{comment}
\begin{table}[!t]\caption{SVC-based QoS thresholds.}\label{tab:SVCQoS}
\centering
%\hspace{-2mm}
%\small
\normalsize
\begin{tabularx}{0.47\textwidth}{|l|l|c|}
\hline \hline 
%\centering
\textbf{Enhancement layers}& \textbf{QoS} & Encoding rate \\
\hline 
0 (base layer) & Low & $0.34$~(\%) \\
1 (base $+$ 1 enhanc. layer) & Medium & $0.62$~(\%) \\
2 (base $+$ 2 enhanc. layers)& High & $0.87$~(\%) \\
\hline \hline
\end{tabularx}
\end{table}
\end{comment}

%\rev{write down some setup you considered for video recording, video encoding (SVC parameters), etc.}

%\rev{the key role of software defined modems here, a little bit about the testbed}

%\rev{M:From the application perspective, we would like to show the accumulation of the litter/density/distribution/number of objects (at least large items) in each patch}
\begin{figure*}[t]
\centering
\setlength{\intextsep}{-2pt}
\par\vspace{\intextsep}
\begin{tabular}{cccc}
\hspace{-0.25in}
\includegraphics[width=0.28\textwidth
]{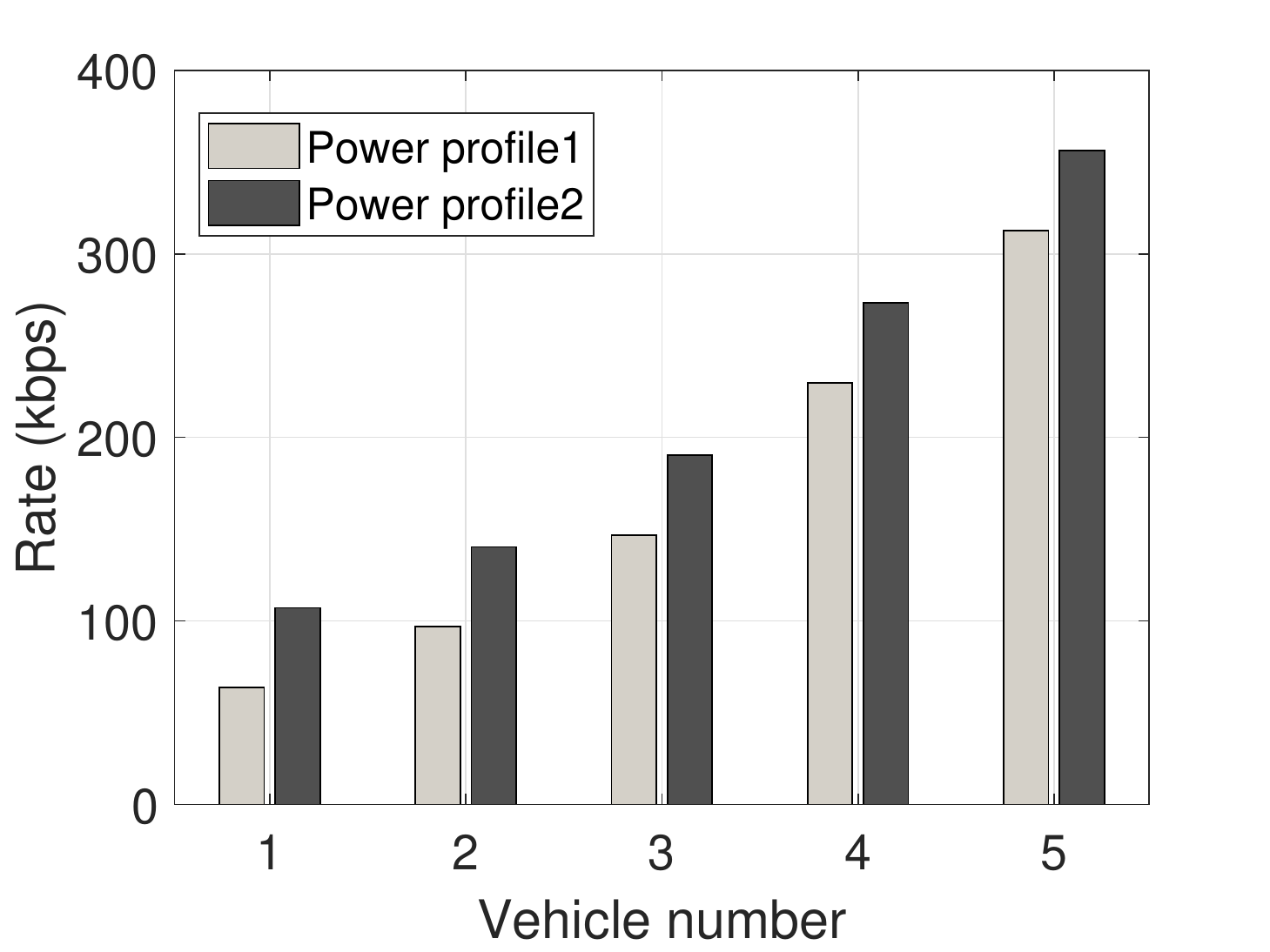} &  
\hspace{-10mm}
\includegraphics[width=0.28\textwidth
]{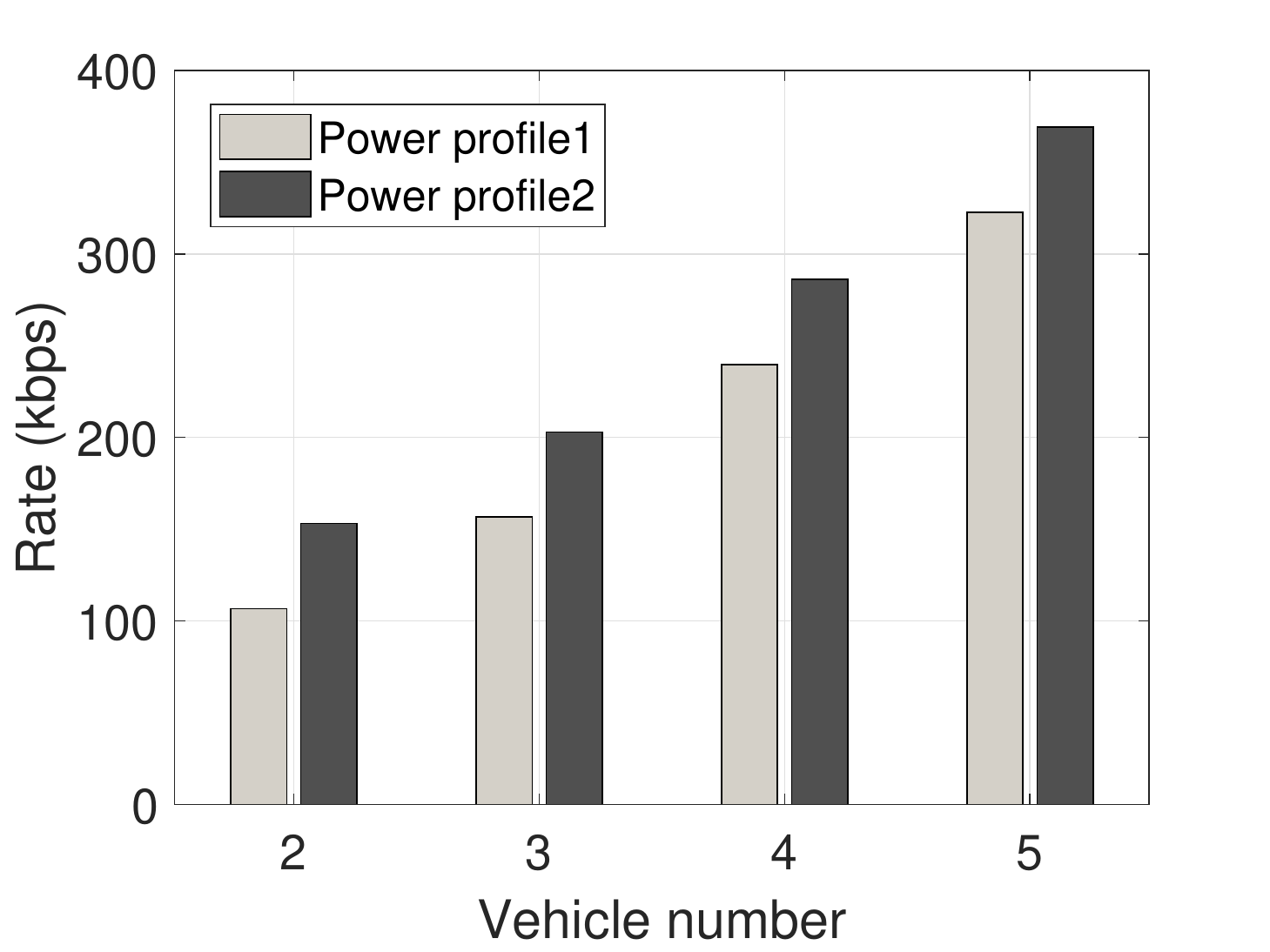} &  
\hspace{-10mm}
\includegraphics[width=0.28\textwidth
]{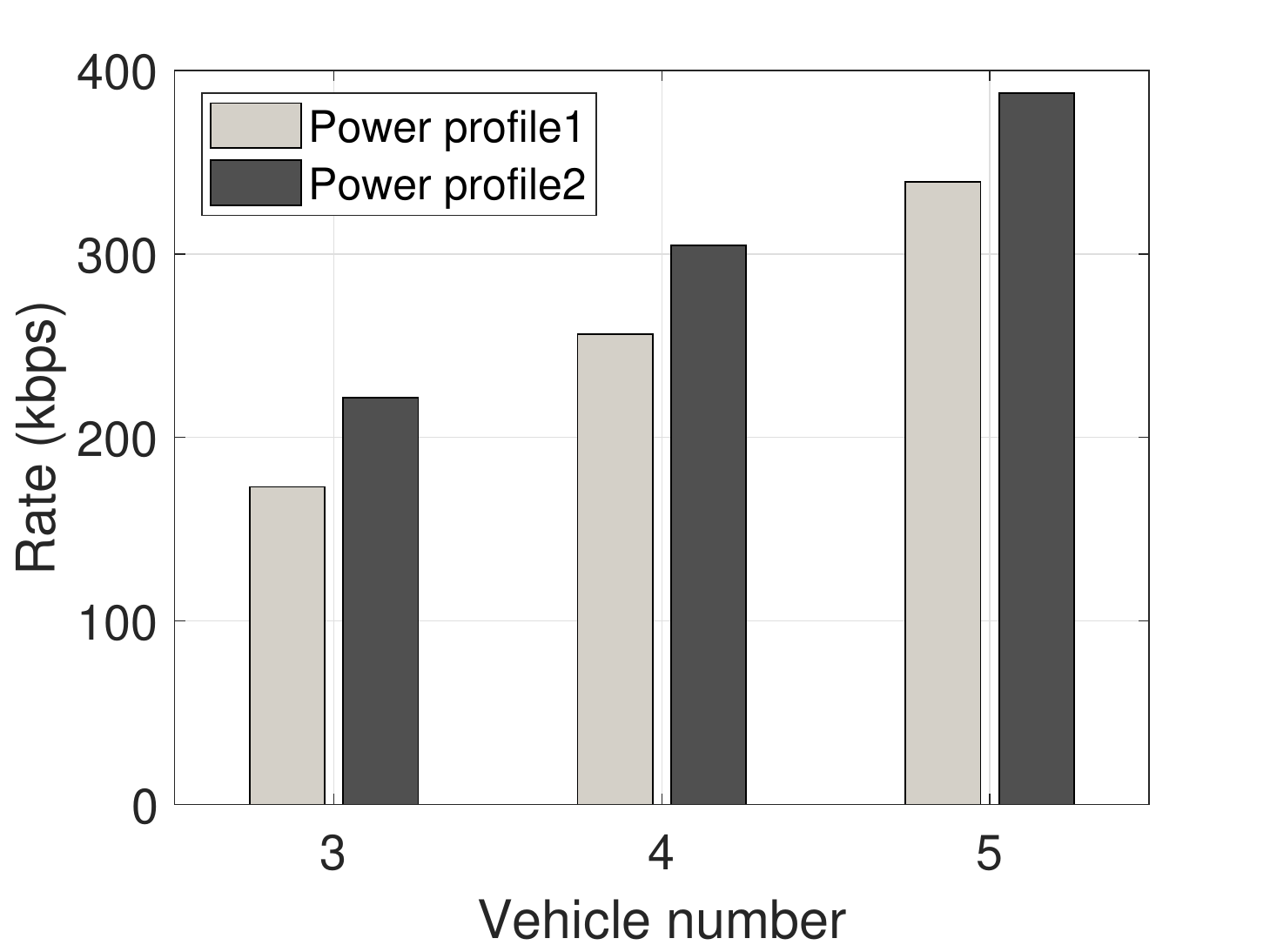} &  
\hspace{-10mm}
\includegraphics[width=0.28\textwidth
]{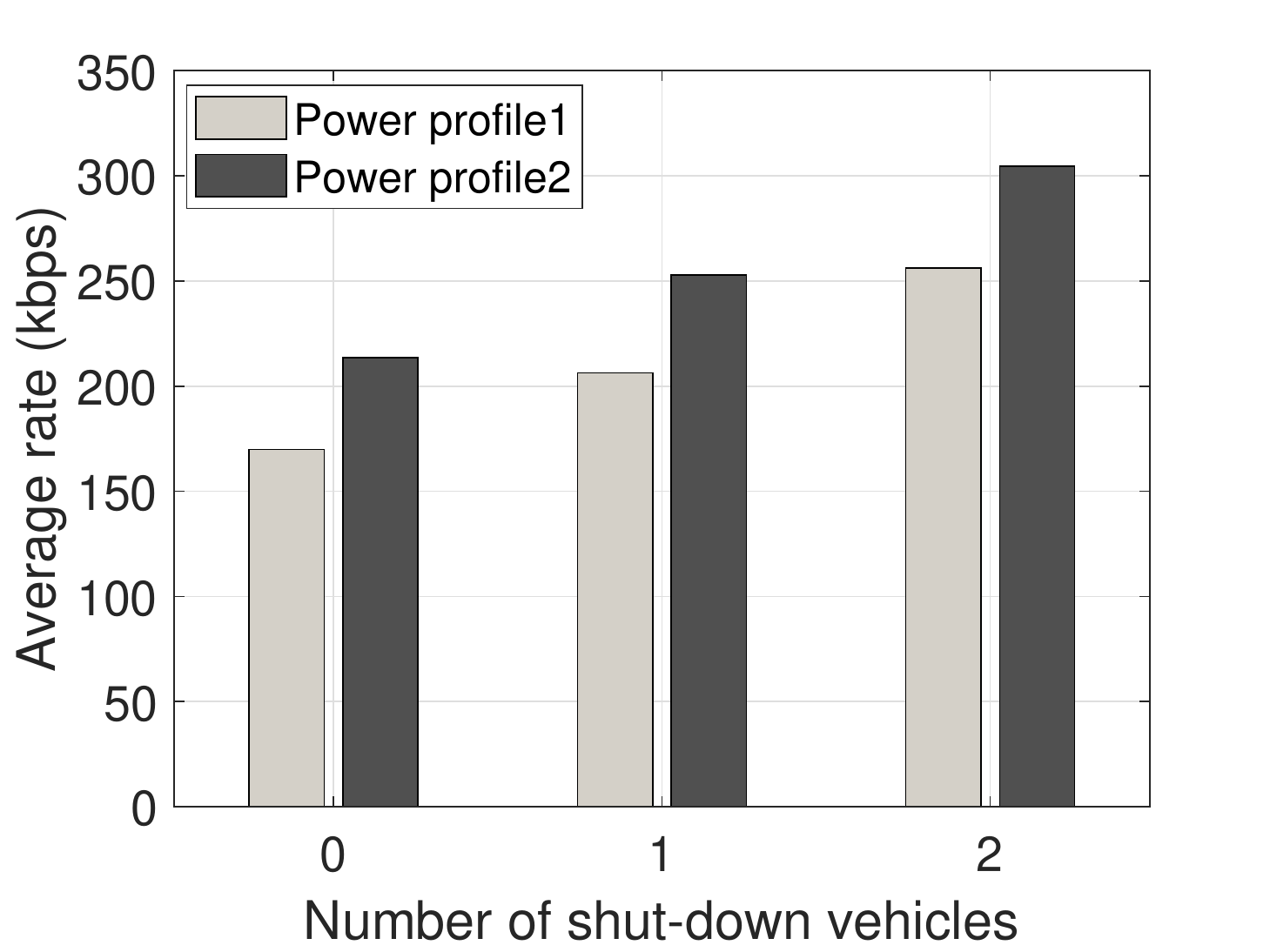}\\  
 \small (a) & \small (b)  & \small (c) & \small (d)
\end{tabular}
\caption{Optimal received rate at different vehicles which are sorted based on their channel quality for two defined power profiles. In (a)~all vehicles are active; (b)~the vehicle with the worst channel quality is shut down; (c)~two vehicles are shut down; (d)~Achieved broadcast rate for two defined power profiles when the number of shut down vehicles changes.}\label{fig:rates_3fig}
%\vspace{-2mm}
\end{figure*}

\begin{comment}
\begin{figure*}[t]
\centering
\setlength{\intextsep}{-2pt}
\par\vspace{\intextsep}
\begin{tabular}{cccc}
\hspace{-0.25in}
\includegraphics[width=0.75\columnwidth
]{fig/rate_5vehicles} &  
\hspace{-11mm}
\includegraphics[width=0.75\columnwidth
]{fig/rate_4vehicles} &  
\hspace{-11mm}
\includegraphics[width=0.75\columnwidth
]{fig/rate_3vehicles} &  
\hspace{-11mm}
\includegraphics[width=0.75\columnwidth
]{fig/rate_3vehicles}\\  
 \small (a) & \small (b)  & \small (c) & \small (c)
\end{tabular}
\caption{Optimal received rate at different vehicles which are sorted based on their channel quality for two defined power profiles. In (a) all vehicles are active; (b) the vehicle with the worst channel quality is shut down; (c) two vehicles are shut down.}\label{fig:rates_3fig}
\vspace{-2mm}
\end{figure*}
\end{comment}

\begin{figure*}[t]
\centering
\begin{tabular}{ccc}
\hspace{-0.25in}
\includegraphics[width=0.74\columnwidth
]{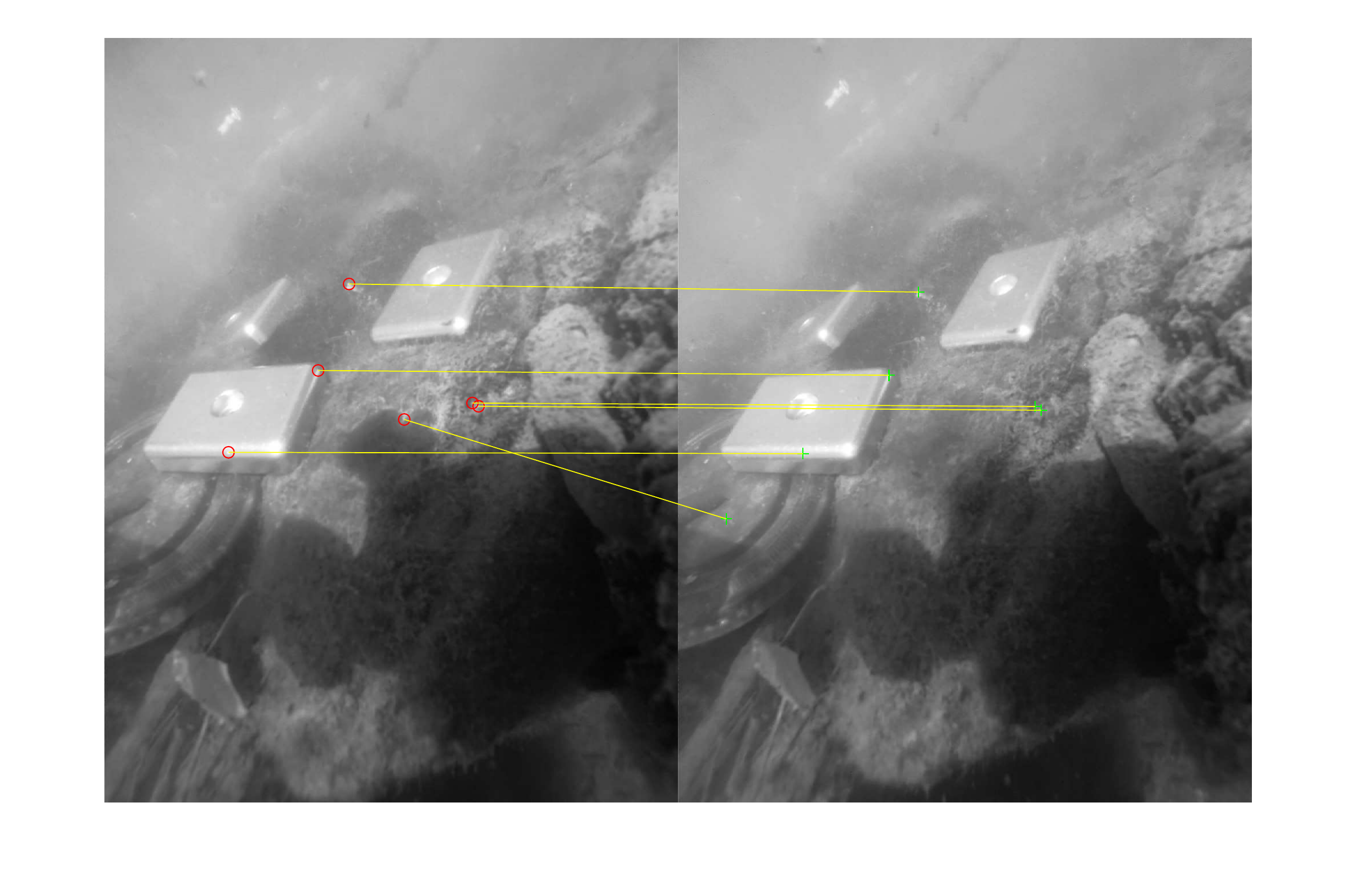} &  
\hspace{-9mm}
\includegraphics[width=0.74\columnwidth
]{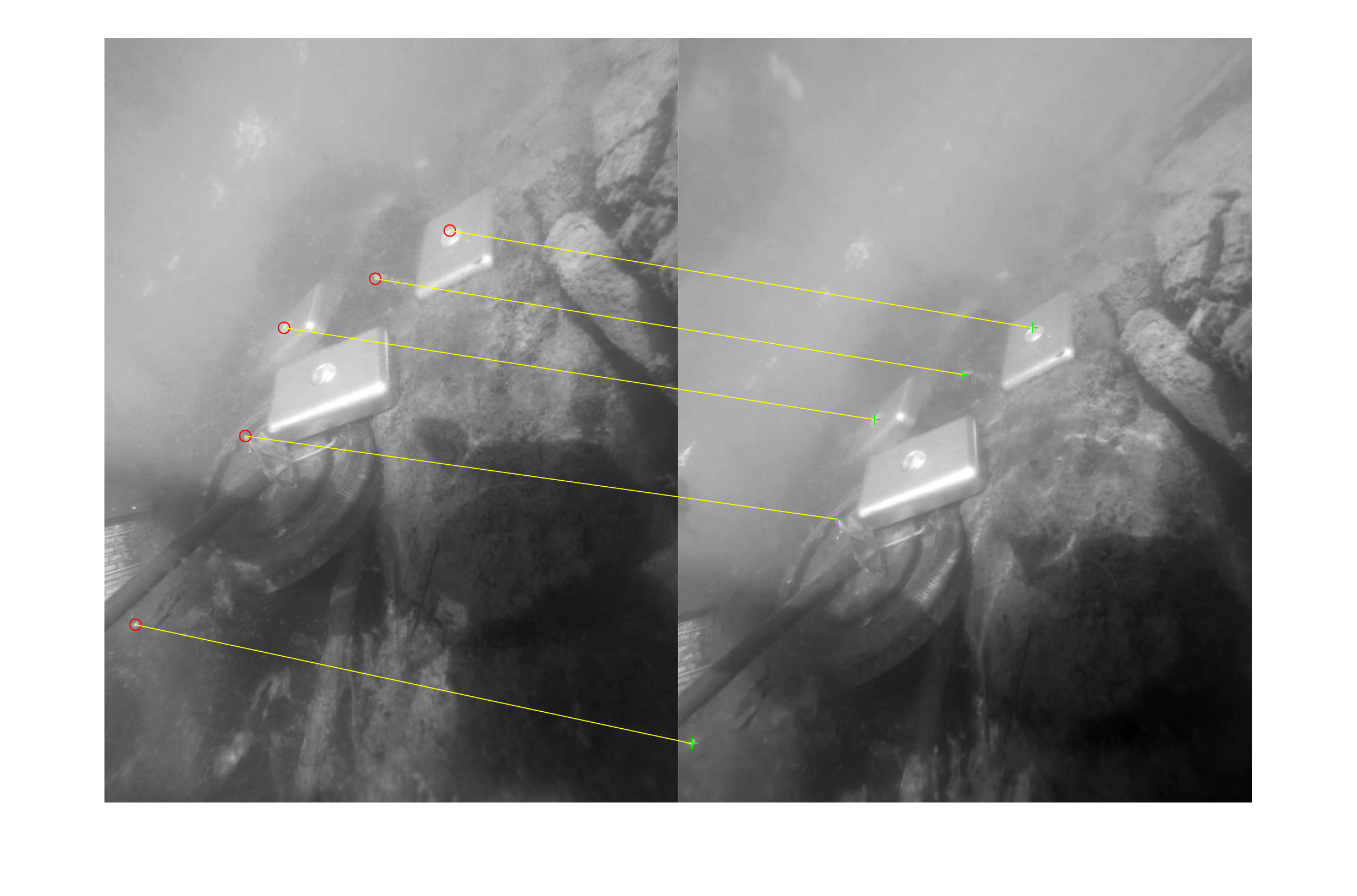} &  
\hspace{-9mm}
\includegraphics[width=0.74\columnwidth
]{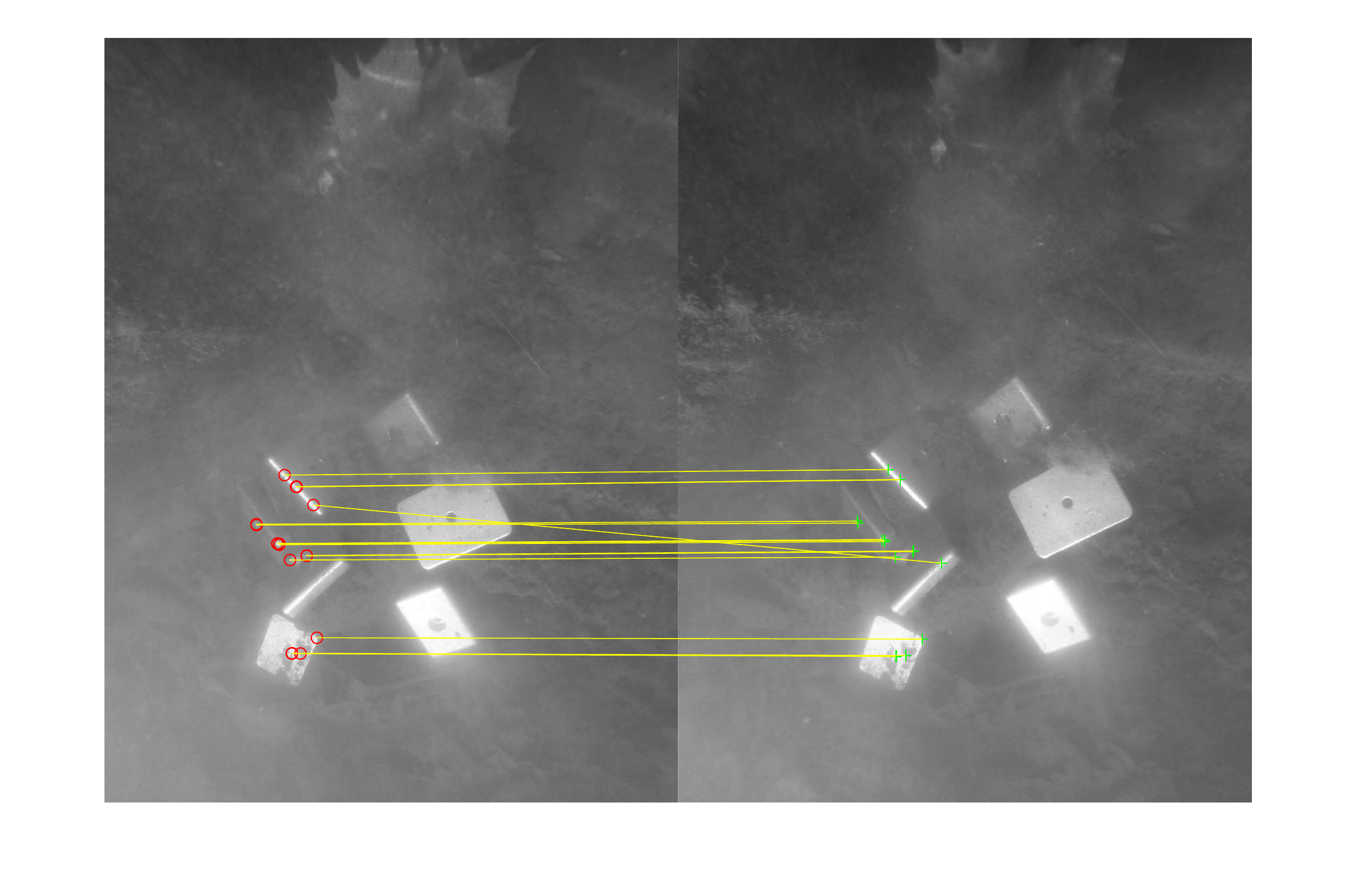} \\  %\vspace{-2mm}
 \small (a) & \small (b)  & \small (c)
\end{tabular}
\caption{Feature matching for different vehicles.}\label{fig:BER_1}
%\vspace{-2mm}
\end{figure*}

\textbf{Results:}
Fig.~\ref{fig:Eval2} shows the effect of the acoustic communication channel on the quality of the received video. The passband channel bandwidth
is $100~\rm{kHz}$ with carrier frequency of $100~\rm{kHz}$ and the sampling rate is $200~\rm{kHz}$. In Figs.~\ref{fig:Eval2}(a)-(c), the original successive frames are shown, while in Figs.~\ref{fig:Eval2}(d)-(f) the quality of the received signal through a good channel is compared to the quality of the received signal through a low to average channel in Figs.~\ref{fig:Eval2}(g)-(i).

Fig.~\ref{fig:svc_layers} depicts different SVC layers of a single selected frame from the captured video. Fig.~\ref{fig:svc_layers}(a) shows the base layer and Fig.~\ref{fig:svc_layers}(b)-(e) represent the base and $1$ to $4$ enhancement layers of the original captured video. The corresponding frame rates for these layers are $1.8750,3.75,7.5,15,30$, respectively with the minimum bit rates of $100.9,179.4,293.3,415.3,517.5~\rm{kbps}$. The corresponding PSNR values are $45.1,44.14,43.31,42.68$ and $42.19~\rm{dB}$, respectively. Fig.~\ref{fig:svc_layers}(f)-(j) show the associated base and enhancement layers of the same frame, when passed through our testbed. Note that the difference between number of enhancement layers can be distinguished better in the video.          

Figs.~\ref{fig:rates_3fig}(a)-(c) demonstrate the optimal received rates at different vehicles as a result of solving the proposed optimization problems. The vehicles are sorted based on their channel quality for two different power profiles. 
In Fig.~\ref{fig:rates_3fig}(a), all the vehicles which are able to receive the base layer video are assumed in active mode. The vehicle which experiences a better channel receives the video with a higher rate. Fig.~\ref{fig:rates_3fig}(b)-(c) shows the vehicles with the worst channel quality are shut down (one vehicle and two vehicles in these two figures, respectively). Fig.~\ref{fig:rates_3fig}(d) represents the proposed solution for the broadcast rate when variable number of vehicles are shut down. Two different power profiles are considered. By shutting down the vehicles with a low channel quality, the average broadcast rate is improved as shown in this figure. However, QoE in the result decreases since less vehicles are involved in the procedure, as explained in the solution.

\begin{figure}[t!]
\centering
\begin{tabular}{cc}
\hspace{-5mm}
\includegraphics [width=0.28\textwidth]{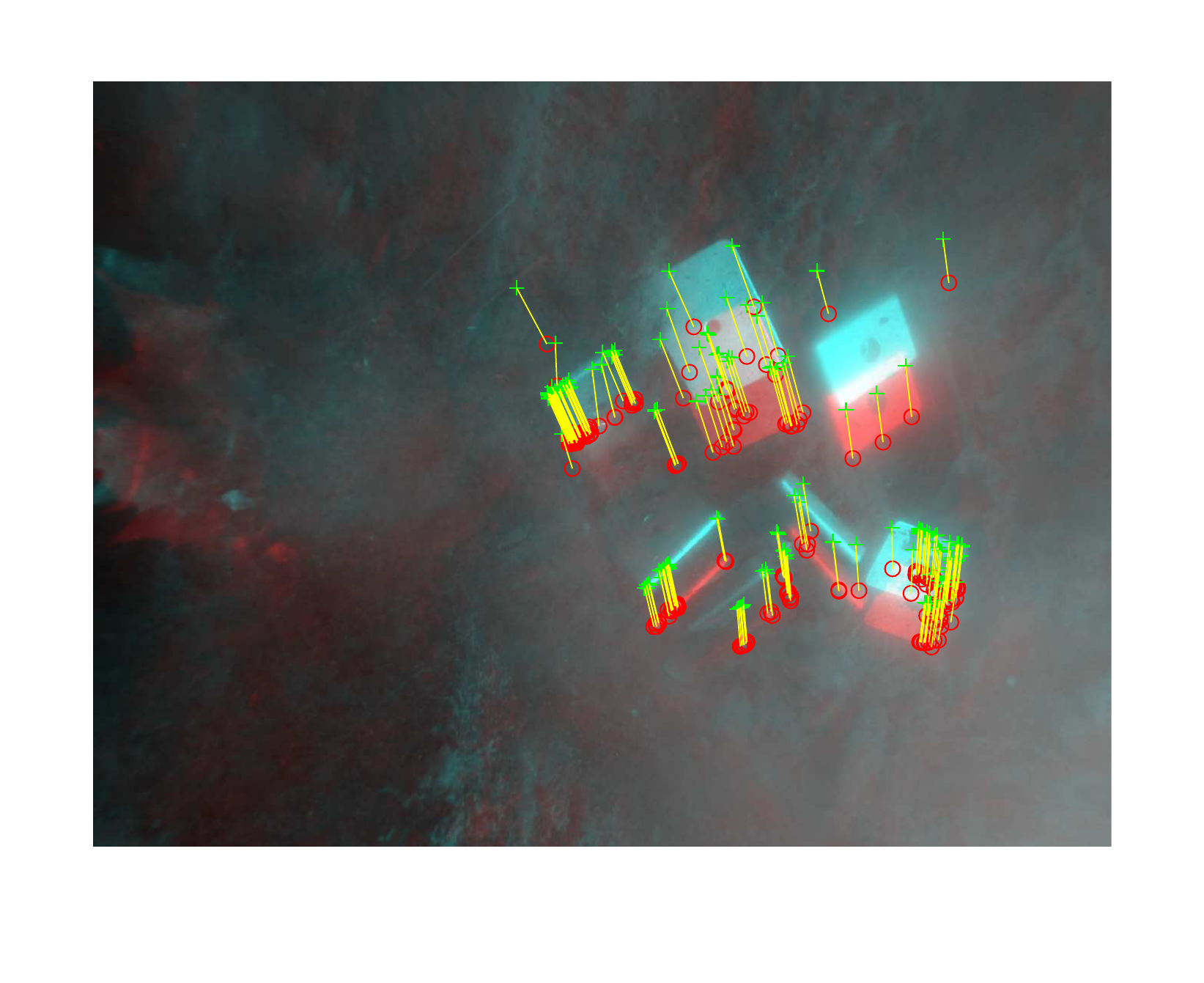} \hspace{-0.8cm}
\includegraphics [width=0.28\textwidth]{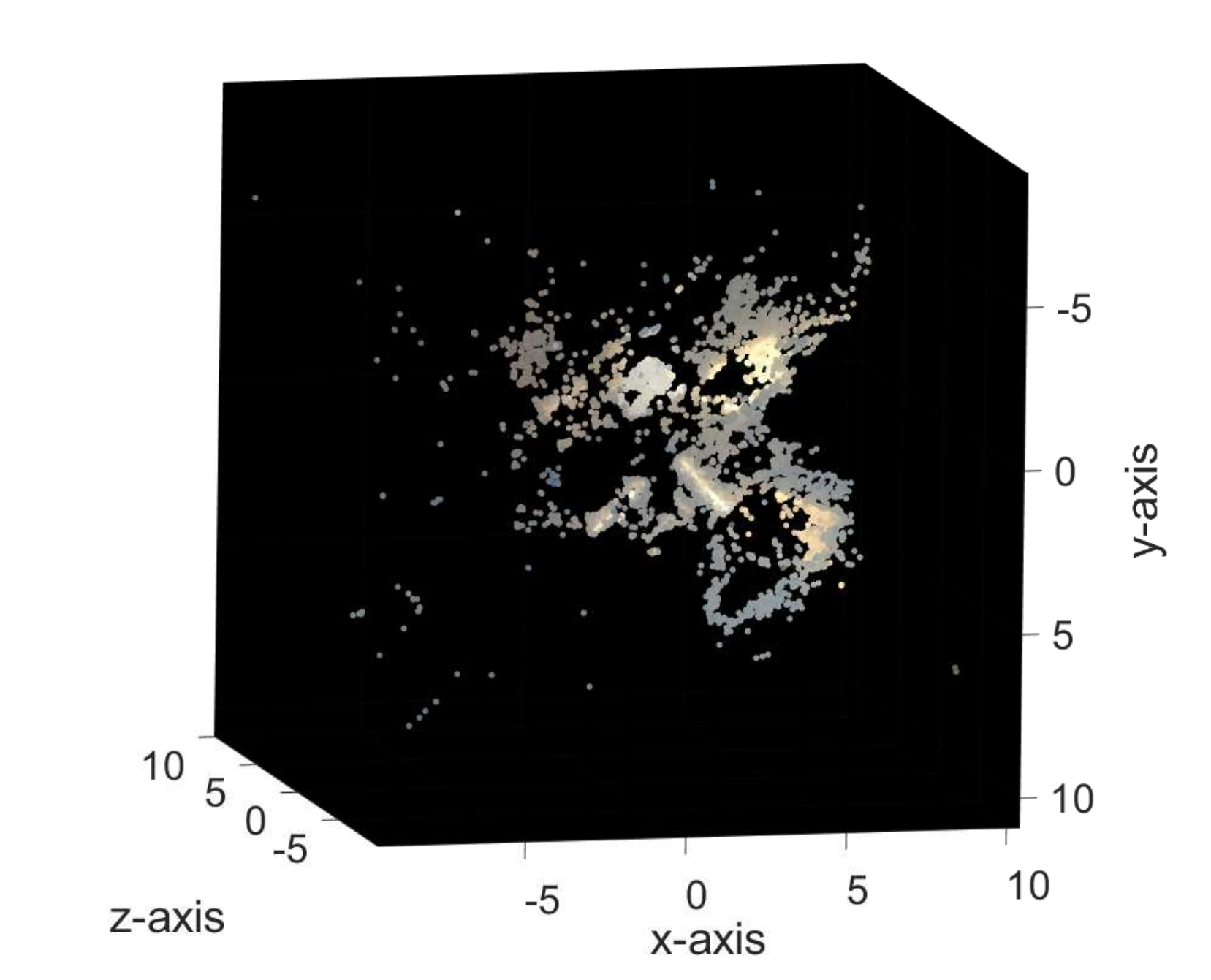}\\
\hspace{0.05cm} (a)   \hspace{3cm} (b) 
\end{tabular}
\caption{(a) Tracked points; (b) Reconstructed map.}\label{fig:mapreconst}
%\vspace{-1mm}
\end{figure}

Figs.~\ref{fig:BER_1}(a)-(c) show the output of the feature matching and reconstruction based on the proposed algorithm. As shown in these figures, each vehicle observes the region of interest partially since there are serious problems with lighting, scattering, turbidity, and clarity when taking underwater videos. In Figs.~\ref{fig:BER_1}(a)-(b), the vehicles detect three objects, while from other perspective, as shown in Fig.~\ref{fig:BER_1}(c), six objects are detected. 
Fig.~\ref{fig:mapreconst} shows the final steps towards map reconstruction. Fig.~\ref{fig:mapreconst}(a) represents the tracked features in the shared images and Fig.~\ref{fig:mapreconst}(b) is the reconstructed map of the region. The map can be used as a QoE metric to evaluate how accurate the desired map should be.     

\begin{comment}
\begin{figure}
\centering 
\setlength{\intextsep}{-2pt}
\par\vspace{\intextsep}
\includegraphics[width=0.37\textwidth]{fig/rate_average}
\caption{Achieved broadcast rate for two defined power profiles when the number of shut down vehicles changes.}\label{fig:rate}
	%\vspace{-5mm}
\end{figure}
\end{comment}

%%%%%%%%%%%%%%%%%%%%%%%%%

\balance

\section{Conclusions and Future Work}\label{sec:Con}
A novel in-network coordination that employed Scalable Video Coding~(SVC) was introduced. Large amounts of data such as videos underwater is not easy to transmit due to the error-prone underwater channel. This paper investigated sharing SVC streams among AUVs in a multicast manner in which the vehicles with different capabilities/channel can be served by a single scalable stream to perform in-network map reconstruction. Performance evaluation was presented based on experiments using video captured from the Raritan River, New Jersey and transmitted through our software-defined acoustic testbed, in addition to simulation. %More experiments are in progress to evaluate the performance of our solution using a software-defined acoustic testbed in the Raritan River, New Jersey. 
In the future, we will extend our current solution to other efficient encoding schemes such as High Efficiency Video Coding~(HEVC) in order to maximize the quality we can achieve under limited bandwidth constraints. Different compression rates will be compared and the effect of lighting, back scattering, and the turbidity will be accounted for and evaluated.

\textbf{Acknowledgements:}
This work was supported by the NSF Award No.~1763964. The authors thank Prince Bose and Zhuoran Qi, Rutgers/ECE graduate students, for their help with the experiments.

\bibliographystyle{IEEEtran}
\bibliography{ref}

%\todo{doublecheck the consistency in the references; also, we can drop a few and get to 30}

\end{document}